\documentclass[11pt, a4paper]{article}
\usepackage{amssymb,here}
\usepackage{graphicx}

\begin{document}

\title{Existence of exact solution of 
the Susceptible-Exposed-Infectious-Recovered 
(SEIR) epidemic model}
\author{Norio Yoshida\thanks{Department of Mathematics, University of Toyama, Toyama, 930-8555 Japan (E-mail: norio.yoshidajp@gmail.com)}}
\date{}

\maketitle
\begin{abstract}
Exact solutions of the SEIR epidemic model are derived, and various properties of solutions 
are obtained directly from the exact solution. 
In this paper Abel differential equations play an important role in 
establishing the exact solution of SEIR differential system, in particular 
the number of infected individuals can be represented in a simple form 
by using a positive 
solution of an Abel differential equation. It is shown that 
the parametric form of the exact solution satisfies 
some linear differential system including a positive solution of an Abel 
differential equation. 
\end{abstract}
\vspace{2ex}
\noindent
{\it Keywords and phrases}: Exact solution, SEIR epidemic model, 
initial value problem, differential system, 
Abel differential equation of the second kind. 
\\

\noindent
{\it 2010 Mathematics Subject Classification}: 34A34

\newtheorem{Df}{Definition}
\newtheorem{thm}{Theorem}
\newtheorem{lem}{Lemma}
\newtheorem{cor}{Corollary}
\newtheorem{ex}{Example}
\newtheorem{rem}{Remark}
\newtheorem{prop}{Proposition}
\newcommand{\qed}{\hfill$\Box$}
\font\bi=cmbxti12 
\newcommand{\Proof}{\rm\vspace{1ex}\noindent{\bi Proof.}}


\section{Introduction}

Recently there has been considerable interest in mathematical approach 
to epidemic models.  Beginning with the pioneering work of 
Bernoulli \cite{b60}, a vast research literature has been published so far 
(cf. \cite{bdw08, c93, f80}), 
and studies of epidemic models have become one of the important areas 
in mathematical epidemiology. 
We mention in particular the paper of Kermack and McKendrick \cite{km27} 
in which the Susceptible-Infectious-Recovered (SIR) model was 
proposed. 

Existence of exact solutions of the epidemic models has been investigated 
in recent years. We refer to Bohner, Streipert and Torres \cite{bst19}, 
Harko, Lobo and Mak \cite{hlm14}, Shabbir, Khan and Sadiq \cite{sks10} 
for SIR epidemic models. 
The Susceptible-Exposed-Infectious-Recovered (SEIR) model has been 
an important subject to study, see, e.g., \cite{ee04, fsaa18, lm95, lgwk99, 
wgk18, whkb20}. 
However, there appears to be no known results about exact solutions of 
SEIR epidemic models. The objective of this paper is to investigate the 
existence of exact solutions of the following SEIR epidemic model: 

\begin{eqnarray}
   \frac{dS(t)}{dt} & = & -\beta S(t)I(t),   \label{ny1} \\
   \frac{dE(t)}{dt} & = & \beta S(t)I(t) - \delta E(t),  \label{ny2} \\
   \frac{dI(t)}{dt} & = & \delta E(t) - \gamma I(t),  \label{ny3} \\
   \frac{dR(t)}{dt} & = & \gamma I(t)  \label{ny4}
\end{eqnarray}
for $t > 0$, where $\beta, \gamma$ and $\delta$ are positive constants. 
The initial condition to be considered is 
the following: 
\begin{equation}
   S(0) = \tilde{S},\ E(0) = \tilde{E},\ I(0) = \tilde{I},\ R(0) = \tilde{R}, 
   \label{ny5}
\end{equation}
where 
$\tilde{S} + \tilde{E} + \tilde{I} + \tilde{R} = N (\mbox{posiive constant})$. 
Since 
$$
   \frac{d}{dt}\left(S(t)+E(t)+I(t)+R(t)\right) 
   = \frac{dS(t)}{dt}+\frac{dE(t)}{dt}+\frac{dI(t)}{dt}+\frac{dR(t)}{dt} = 0 
$$
by (\ref{ny1})--(\ref{ny4}), it follows that 
$$
   S(t) + E(t) + I(t) + R(t) = k \ (t \geq 0) 
$$
for some constant $k$.  In view of the fact that 
$$
   k = S(0) + E(0) + I(0) + R(0) = \tilde{S}+\tilde{E}+\tilde{I}+\tilde{R}=N,
$$
we conclude that 
$$
   S(t) + E(t) + I(t) + R(t) = N\ (t \geq 0). 
$$
It is assumed throughout this paper that:
\begin{itemize}
   \item[(A$_{1}$)] $\tilde{I} > 0$; 
   \item[(A$_{2}$)] $\displaystyle\tilde{E} > \frac{\gamma}{\delta}\tilde{I}$; 
   \item[(A$_{3}$)] $\displaystyle \tilde{S} > 
                     \frac{\delta \tilde{E}}{\beta \tilde{I}}$; 
   \item[(A$_{4}$)] $\tilde{R} \geq 0$ satisfies
$$
   N > \tilde{S}e^{(\beta/\gamma)\tilde{R}} + \tilde{R}. 
$$
\end{itemize}

In Section 2 we show that the exact solution of the initial value problem 
(\ref{ny1})--(\ref{ny5}) can be represented in a parametric form. 
In Section 3 we discuss the global existence and uniqueness of solutions of 
an initial value problem for an Abel differential equation of the 
second kind, and in Section 4 we derive the exact solution of 
the initial value problem (\ref{ny1})--(\ref{ny5}) and we show that 
the parametric solution in Section 2 can be obtained by 
solving an initial value problem for some linear differential system 
where a positive solution of an Abel differential equation appears in 
the coefficients. 
Section 5 is devoted to various properties 
of solutions of SEIR epidemic model.


\section{Exact solution of an initial value problem for SEIR 
differential system (I)} 

First we need the following important lemma. 

\begin{lem} \label{ny:lem1} 
If $S(t) > 0$ for $t > 0$, then the following holds{\rm :} 
\begin{equation}
   R''(t) + (\gamma+\delta)R'(t) 
   = \gamma\delta \left(N - \tilde{S}e^{(\beta/\gamma)\tilde{R}}
     e^{-(\beta/\gamma)R(t)} - R(t) \right)\ (t > 0). 
   \label{ny6}
\end{equation}
\end{lem}

{\Proof}
It follows from (\ref{ny1}) and (\ref{ny4}) that
$$
   R'(t) = \gamma I(t) = \gamma\left(\frac{S'(t)}{-\beta S(t)}\right) 
         = - \frac{\gamma}{\beta}\bigl(\log S(t)\bigr)', 
$$
and integrating the above on $[0,t]$ yields 
$$
   R(t) - \tilde{R} 
   = - \frac{\gamma}{\beta}\bigl( \log S(t) - \log \tilde{S}\bigr). 
$$
Therefore we obtain 
$$
   \log S(t) = - \frac{\beta}{\gamma}\bigl(R(t) - \tilde{R}\bigr) 
               + \log \tilde{S} 
$$
and hence
\begin{equation}
   S(t) = \exp \left(\log \tilde{S} - \frac{\beta}{\gamma}R(t) 
                     + \frac{\beta}{\gamma} \tilde{R} \right) 
        = \tilde{S}e^{(\beta/\gamma)\tilde{R}} e^{-(\beta/\gamma)R(t)}. 
          \label{ny7} 
\end{equation}
From (\ref{ny4}) we see that $I(t) = R'(t)/\gamma$, and hence 
$I'(t) = R''(t)/\gamma$. Therefore, it follows from (\ref{ny3}) that 
\begin{equation}
   E(t) = \frac{1}{\delta} \bigl(I'(t) + \gamma I(t)\bigr) 
     = \frac{1}{\delta} \left(\frac{R''(t)}{\gamma} + R'(t)\right). 
       \label{ny8}
\end{equation}
Taking account of (\ref{ny4}), (\ref{ny7}) and (\ref{ny8}), we get 
\begin{eqnarray*}
   \frac{R'(t)}{\gamma}
   & = & I(t) \\
   & = & N - S(t) - R(t) - E(t) \\
   & = & N - \tilde{S}e^{(\beta/\gamma)\tilde{R}}e^{-(\beta/\gamma)R(t)} 
         - R(t) - \frac{1}{\delta} \left(\frac{R''(t)}{\gamma} + R'(t)\right) 
\end{eqnarray*}
which implies 
$$
   \frac{1}{\gamma\delta}R''(t) 
   = N - \tilde{S}e^{(\beta/\gamma)\tilde{R}}e^{-(\beta/\gamma)R(t)} - R(t) 
     - \left(\frac{1}{\delta} + \frac{1}{\gamma}\right)R'(t). 
$$
Multiplying the above by $\gamma\delta$ yields the desired 
identity (\ref{ny6}). 
\qed

By a {\it solution} of the SEIR differential system (\ref{ny1})--(\ref{ny4}) 
we mean a vector-valued function $(S(t),E(t),I(t),R(t))$ of class $C^{1}(0,\infty) \cap C[0,\infty)$ which satisfies (\ref{ny1})--(\ref{ny4}). 
Associated with every continuous function $f(t)$ on $[0,\infty)$, we define
$$
   f(\infty) := \lim_{t \to \infty} f(t).
$$

\begin{lem} \label{ny:lem2} 
Let $(S(t),E(t),I(t),R(t))$ be a solution of the 
SEIR differential system {\rm (\ref{ny1})--(\ref{ny4})} 
such that $S(t) > 0$, $E(t) > 0$ and $I(t) > 0$ for $t > 0$. 
Then there exists the limit $R(\infty)$. 
\end{lem}

{\Proof}
Since $R'(t) = \gamma I(t) > 0\ (t > 0)$ and 
$$
   R(t) = N - S(t) - E(t) - I(t) < N \ (t > 0), 
$$
we find that $R(t)$ is increasing on $[0,\infty)$ and bounded from above. Hence there exists the limit $R(\infty)$.
\qed

\begin{thm} \label{ny:thm1}
Let $(S(t),E(t),I(t),R(t))$ be a solution of the initial value problem 
{\rm (\ref{ny1})--(\ref{ny5})} such that $S(t) > 0$, $E(t) > 0$ and 
$I(t) > 0$ for $t > 0$. 
Then the solution $(S(t),E(t),I(t),R(t))$ can be 
represented in the following parametric form{\rm :} 
\begin{eqnarray}
   S(\varphi(u)) & = & \tilde{S}e^{(\beta/\gamma)\tilde{R}}u, \label{ny9} \\
   E(\varphi(u)) & = & \tilde{E}e^{-\delta\varphi(u)} 
           + \tilde{S}e^{(\beta/\gamma)\tilde{R}}e^{-\delta\varphi(u)}
           \int_{u}^{e^{-(\beta/\gamma)\tilde{R}}}
           e^{\delta\varphi(v)}dv, \label{ny10} \\
   I(\varphi(u)) & = & N - \tilde{S}e^{(\beta/\gamma)\tilde{R}}u 
           + \frac{\gamma}{\beta}\log u 
           - \tilde{E}e^{-\delta\varphi(u)}  \nonumber\\
     &   &  \qquad - \tilde{S}e^{(\beta/\gamma)\tilde{R}}e^{-\delta\varphi(u)}
           \int_{u}^{e^{-(\beta/\gamma)\tilde{R}}}
           e^{\delta\varphi(v)}dv,   \label{ny11} \\
   R(\varphi(u)) & = & - \frac{\gamma}{\beta}\,\log u \label{ny12}
\end{eqnarray}
for $e^{-(\beta/\gamma)R(\infty)} < u \leq e^{-(\beta/\gamma)\tilde{R}}$, 
where 
\begin{equation}
   t = \varphi(u) = \int_{u}^{e^{-(\beta/\gamma)\tilde{R}}}
                     \frac{d\xi}{\xi\psi(\xi)},  \label{ny13}
\end{equation}
with $\psi(u)$ satisfying the Abel differential equation of the second kind 
\begin{equation}
   \psi'\psi - \frac{\gamma+\delta}{u}\psi 
   = - \delta\,\frac{\beta N - \beta \tilde{S}e^{(\beta/\gamma)\tilde{R}}u
     + \gamma \log u}{u}, 
     u \in (e^{-(\beta/\gamma)R(\infty)}, e^{-(\beta/\gamma)\tilde{R}}),   
     \label{ny14}
\end{equation}
and the following conditions
\begin{eqnarray*}
   & & \psi\bigl(e^{-(\beta/\gamma)\tilde{R}}\bigr) = \beta \tilde{I}, \\
   & & \lim_{u \to e^{-(\beta/\gamma)R(\infty)}+0} \psi(u) = 0, \\
   & & \psi(u) > 0\ {\rm in}\ 
       (e^{-(\beta/\gamma)R(\infty)}, e^{-(\beta/\gamma)\tilde{R}}]. 
\end{eqnarray*}
\end{thm}

{\Proof}
Since $R'(t) = \gamma I(t) > 0$ for $t > 0$ by means of (\ref{ny4}), 
we see that $R(t)$ is increasing on $[0,\infty)$. 
We note that there exists the limit $R(\infty)$ by Lemma \ref{ny:lem2}. 
Then 
$u = u(t) = e^{-(\beta/\gamma)R(t)}$ is decreasing on $[0,\infty)$, 
$e^{-(\beta/\gamma)R(\infty)} < u \leq e^{-(\beta/\gamma)\tilde{R}}$, 
and $\lim_{t \to \infty} u(t) = e^{-(\beta/\gamma)R(\infty)}$. 
Hence there exists the inverse function 
$\varphi(u) \in 
C^{1}(e^{-(\beta/\gamma)R(\infty)},e^{-(\beta/\gamma)\tilde{R}})$ 
of $u = u(t)$ 
such that 
$$
   t = \varphi(u) \quad
   \left(e^{-(\beta/\gamma)R(\infty)} < u \leq 
   e^{-(\beta/\gamma)\tilde{R}}\right), 
$$
$\varphi(u)$ is decreasing in 
$(e^{-(\beta/\gamma)R(\infty)}, e^{-(\beta/\gamma)\tilde{R}}]$, 
$\lim_{u \to e^{-(\beta/\gamma)R(\infty)}+0} \varphi(u) = \infty$, 
and $\varphi\bigl(e^{-(\beta/\gamma)\tilde{R}}\bigr) = 0$. 
Substituting $t = \varphi(u)$ into (\ref{ny6}) in Lemma \ref{ny:lem1} 
yields
\begin{equation}
      R''(\varphi(u)) + (\gamma+\delta)R'(\varphi(u)) 
   = \gamma\delta \left(N - \tilde{S}e^{(\beta/\gamma)\tilde{R}}
     e^{-(\beta/\gamma)R(\varphi(u))} - R(\varphi(u)) \right)
   \label{ny15}
\end{equation}
for $e^{-(\beta/\gamma)R(\infty)} < u < e^{-(\beta/\gamma)\tilde{R}}$. 
Differentiating both sides of 
$u = e^{-(\beta/\gamma)R(\varphi(u))}$ 
with respect to $u$, we obtain 
\begin{eqnarray*}
   1 & = & -\frac{\beta}{\gamma}R'(\varphi(u))\varphi'(u)
           e^{-(\beta/\gamma)R(\varphi(u))} \\
     & = & -\frac{\beta}{\gamma}R'(\varphi(u))\varphi'(u)u, 
\end{eqnarray*}
and therefore 
\begin{equation}
   R'(\varphi(u)) = -\frac{\gamma}{\beta} \frac{1}{\varphi'(u)u}. 
   \label{ny16}
\end{equation}
Since $R'(t) \in C^{1}(0,\infty)$ by means of (\ref{ny4}) and 
$\varphi(u) \in 
C^{1}(e^{-(\beta/\gamma)R(\infty)},e^{-(\beta/\gamma)\tilde{R}})$, 
we see that $R'(\varphi(u)) \in 
C^{1}(e^{-(\beta/\gamma)R(\infty)},e^{-(\beta/\gamma)\tilde{R}})$, 
and consequently $1/(\varphi'(u)u)$ $\in$ 
$C^{1}(e^{-(\beta/\gamma)R(\infty)},e^{-(\beta/\gamma)\tilde{R}})$. 
We differentiate (\ref{ny16}) with respect to $u$ to get 
$$
   R''(\varphi(u))\varphi'(u) = 
   -\frac{\gamma}{\beta} \left(\frac{1}{\varphi'(u)u}\right)', 
$$
and hence
\begin{equation}
   R''(\varphi(u)) = 
   -\frac{\gamma}{\beta} \left(\frac{1}{\varphi'(u)u}\right)'
   \frac{1}{\varphi'(u)}. 
       \label{ny17}
\end{equation}
It is clear that 
\begin{equation}
   R(\varphi(u)) = - \frac{\gamma}{\beta} \log u  \label{ny18}
\end{equation}
in light of $u = e^{-(\beta/\gamma)R(\varphi(u))}$. 
Combining (\ref{ny15})--(\ref{ny18}) yields 
\begin{eqnarray*}
   &   & -\frac{\gamma}{\beta} \left(\frac{1}{\varphi'(u)u}\right)'
         \frac{1}{\varphi'(u)}
         + (\gamma+\delta)\left( -\frac{\gamma}{\beta} \frac{1}{\varphi'(u)u}
         \right) \\
   & = & \gamma\delta \left(N - \tilde{S}e^{(\beta/\gamma)\tilde{R}}u 
         + \frac{\gamma}{\beta} \log u \right) 
\end{eqnarray*}
or
\begin{eqnarray}
   &   & \frac{\gamma}{\beta} \left(- \frac{1}{\varphi'(u)u}\right)'
         \left( - \frac{1}{\varphi'(u)u}\right)
         - \frac{\gamma}{\beta} \frac{\gamma+\delta}{u}
         \left( - \frac{1}{\varphi'(u)u}
         \right)   \nonumber\\
   & = & - \gamma\delta \frac{1}{u}
         \left(N - \tilde{S}e^{(\beta/\gamma)\tilde{R}}u 
         + \frac{\gamma}{\beta} \log u \right).  \label{ny19}
\end{eqnarray}
Letting 
\begin{equation}
   \psi(u) := - \frac{1}{\varphi'(u)u},  \label{ny20}
\end{equation}
we observe that $\psi(u)$ satisfies (\ref{ny14}). 
Noting that $t = \varphi(u) > 0$ for 
$e^{-(\beta/\gamma)R(\infty)} < u < e^{-(\beta/\gamma)\tilde{R}}$, 
we see from (\ref{ny4}), (\ref{ny16}) and (\ref{ny20}) that
$$
   \psi(u) = \frac{\beta}{\gamma} R'(\varphi(u)) 
   = \beta I(\varphi(u)) > 0
$$
in $(e^{-(\beta/\gamma)R(\infty)}, e^{-(\beta/\gamma)\tilde{R}})$. 
If we define 
\begin{eqnarray*}
   \psi\bigl(e^{-(\beta/\gamma)\tilde{R}}\bigr) 
   & := & \lim_{u \to e^{-(\beta/\gamma)\tilde{R}}-0} \psi(u) 
          = \frac{\beta}{\gamma} 
          \lim_{u \to e^{-(\beta/\gamma)\tilde{R}}-0} R'(\varphi(u)) \\
   & =  & \frac{\beta}{\gamma} \lim_{t \to +0} R'(t) 
          = \frac{\beta}{\gamma} \gamma I(0) 
          = \beta \tilde{I} > 0, 
\end{eqnarray*}
then $\psi(u)$ is a positive continuous function in 
$(e^{-(\beta/\gamma)R(\infty)}, e^{-(\beta/\gamma)\tilde{R}}]$. 
It follows from (\ref{ny20}) that 
$$
   t = \varphi(u) = \int_{e^{-(\beta/\gamma)\tilde{R}}}^{u} 
                \varphi'(\xi)d\xi 
              = \int_{u}^{e^{-(\beta/\gamma)\tilde{R}}}
                     \frac{d\xi}{\xi\psi(\xi)}, 
$$
and therefore (\ref{ny13}) holds. 
Since $\lim_{u \to e^{-(\beta/\gamma)R(\infty)}+0} \varphi(u) = \infty$, 
it is necessary that 
$\lim_{u \to e^{-(\beta/\gamma)R(\infty)}+0} \psi(u) = 0$. 

Now we derive the representation formulae (\ref{ny9})--(\ref{ny12}). 
It follows from (\ref{ny7}) and (\ref{ny18}) that 
\begin{eqnarray*}
   S(\varphi(u)) & = & 
   \tilde{S}e^{(\beta/\gamma)\tilde{R}}e^{-(\beta/\gamma)R(\varphi(u))} 
   = \tilde{S}e^{(\beta/\gamma)\tilde{R}}u, \\
   R(\varphi(u)) & = & - \frac{\gamma}{\beta} \log u,
\end{eqnarray*}
which are the desired representations (\ref{ny9}) and (\ref{ny12}). 
Combining (\ref{ny1}) and (\ref{ny2}), we obtain the first order 
linear differential equation
$$
   E'(t) + \delta E(t) = - S'(t) 
$$
which implies
\begin{equation}
   E(t) = \tilde{E}e^{-\delta t} 
          - e^{-\delta t}\int_{0}^{t} e^{\delta \xi}S'(\xi)d\xi. 
          \label{ny21}
\end{equation}
Differentiating (\ref{ny7}) yields 
\begin{equation}
   S'(t) = - \frac{\beta}{\gamma}\tilde{S}e^{(\beta/\gamma)\tilde{R}}
           R'(t)e^{-(\beta/\gamma)R(t)}. 
           \label{ny22}
\end{equation}
Substitution of (\ref{ny22}) into (\ref{ny21}) gives 
\begin{equation}
   E(t) = \tilde{E}e^{-\delta t} 
          + \frac{\beta}{\gamma}\tilde{S}e^{(\beta/\gamma)\tilde{R}}
          e^{-\delta t} \int_{0}^{t} e^{\delta \xi}R'(\xi)
          e^{-(\beta/\gamma)R(\xi)} d\xi. 
          \label{ny23}
\end{equation}
By changing the variables $R(\xi) = s$, we obtain 
\begin{eqnarray*}
   \Omega := \int_{0}^{t} e^{\delta \xi}R'(\xi)e^{-(\beta/\gamma)R(\xi)} d\xi      & = & \int_{\tilde{R}}^{R(t)} e^{\delta R^{-1}(s)}e^{-(\beta/\gamma)s} ds \\
   & = & \int_{\tilde{R}}^{R(t)} 
         e^{\delta \varphi(e^{-(\beta/\gamma)s})} e^{-(\beta/\gamma)s} ds \\
   & = & \frac{\gamma}{\beta} \int_{R(t)}^{\tilde{R}}
         e^{\delta \varphi(e^{-(\beta/\gamma)s})} 
         \left(e^{-(\beta/\gamma)s}\right)'ds 
\end{eqnarray*}
in light of $R^{-1}(s) = \varphi(e^{-(\beta/\gamma)s})$. 
Letting $v = e^{-(\beta/\gamma)s}$ yields 
\begin{equation}
   \Omega = \frac{\gamma}{\beta} 
   \int_{e^{-(\beta/\gamma)R(t)}}^{e^{-(\beta/\gamma)\tilde{R}}}
         e^{\delta \varphi(v)} dv. 
         \label{ny24}
\end{equation}
Combining (\ref{ny23}) with (\ref{ny24}), we are led to 
\begin{equation}
   E(t) = \tilde{E}e^{-\delta t} 
          + \tilde{S}e^{(\beta/\gamma)\tilde{R}} e^{-\delta t}
          \int_{e^{-(\beta/\gamma)R(t)}}^{e^{-(\beta/\gamma)\tilde{R}}}
          e^{\delta \varphi(v)} dv. 
          \label{ny25}
\end{equation}
Substituting $t = \varphi(u)$ into (\ref{ny25}), we arrive at (\ref{ny10}). 
Since $I(\varphi(u)) = N - S(\varphi(u)) - R(\varphi(u)) - E(\varphi(u))$, 
(\ref{ny11}) follows from (\ref{ny9}), (\ref{ny10}) and (\ref{ny12}). 
\qed

\begin{rem} \label{ny:rem1} \rm
If $\psi(u)$ is a positive solution of (\ref{ny14}), 
then $\Psi(u) = 1/\psi(u)$ is a positive 
solution of the Abel differential equation of the first kind 
$$
   \Psi' = - \frac{\gamma+\delta}{u}\,\Psi^{2} 
           + \delta \frac{\beta N - \beta \tilde{S}e^{(\beta/\gamma)\tilde{R}}u
           + \gamma \log u}{u}\,\Psi^{3}. 
$$
\end{rem}


\section{Abel differential equation of the second kind} 

We study the existence and uniqueness of solutions of an initial value 
problem for the Abel differential equation of the second kind 
\begin{equation}
   \bigl(w' + K \bigr)w = f(x), \quad x_{0} < x < \alpha, 
   \label{ny26}
\end{equation} 
where $K$ is a positive constant and $f(x)$ is a positive continuous function 
in $[x_{0}, \alpha)$ such that 
$f(x)$ is decreasing in $[x_{0}, \alpha)$ or $\lim_{x \to \alpha-0} f(x) = 0$. 
We refer to Abel \cite{a65} and Davis \cite{d62} for Abel differential 
equations. 

If we show that there exists a unique positive solution of the 
initial value problem for 
\begin{equation}
   w' + K = \frac{f(x)}{w}, \quad x_{0} < x < \alpha,  
   \label{ny27}
\end{equation}
subject to the initial condition 
\begin{equation}
   w(x_{0}) = w_{0} > 0, 
   \label{ny28}
\end{equation}
then there exists a unique positive solution of the initial value 
problem (\ref{ny26}), (\ref{ny28}). 

\begin{thm} \label{ny:thm2}
There exists a unique local {\rm ({\it positive})} solution $w(x)$ of 
the initial value problem 
{\rm (\ref{ny27}), (\ref{ny28})}. 
\end{thm}

{\Proof} 
We define the rectangle $\mathcal{R}$ by 
$$
   \mathcal{R} := \left\{ (x,w);\ \vert x-x_{0}\vert \leq \sigma,\ 
                  \vert w-w_{0}\vert \leq \rho \right\}, 
$$
where $0 < \sigma < \alpha - x_{0}$ and $0 < \rho < w_{0}$, and let 
$g(x) \in C[x_{0}-\sigma, x_{0}+\sigma]$ satisfy 
$g(x) = f(x)$ on $[x_{0}, x_{0}+\sigma]$. 
Then we see that
$$
   F(x,w) := - K + \frac{g(x)}{w} \in C(\mathcal{R}) 
$$
and that $F(x,w)$ satisfies a Lipschitz condition with the 
Lipschitz constant 
$$
   L = \frac{\max_{\vert x-x_{0}\vert \leq \sigma} \vert g(x)\vert}
            {(w_{0}- \rho)^{2}}
$$
in view of the fact that
\begin{eqnarray*}
   \left\vert F(x,w_{1}) - F(x,w_{2}) \right\vert 
   & = & \vert g(x)\vert 
       \left\vert \frac{1}{w_{1}} - \frac{1}{w_{2}} \right\vert 
       = \frac{\vert g(x)\vert}{\vert w_{1}w_{2}\vert} \vert w_{1} - w_{2}\vert \\
   & \leq & \frac{\max_{\vert x-x_{0}\vert \leq \sigma} \vert g(x)\vert}
            {(w_{0}- \rho)^{2}} \vert w_{1} - w_{2}\vert 
\end{eqnarray*}
for any $(x,w_{1}), (x,w_{2}) \in \mathcal{R}$. 
Therefore, there exists a unique (positive) solution $\hat{w}(x)$ of 
$w' = F(x,w)$ on $\vert x-x_{0}\vert \leq \tilde{\sigma}$ for which 
$0 < \tilde{\sigma} < \sigma$ and $\hat{w}(x_{0}) = w_{0}$ 
(see, e.g., Coddington and Levinson \cite[Theorem 2.3]{cl55}). 
Then we observe that $\hat{w}(x)$ is a unique (positive) solution of 
$w' = -K + (f(x)/w)$ on $x_{0} \leq x \leq x_{0}+\tilde{\sigma}$ such that 
$\hat{w}(x_{0}) = w_{0}$. 
\qed

We consider the initial value problem for the Abel differential equation 
of the second kind 
\begin{equation}
   y' = - K + \frac{C}{y}, \quad x > x_{0},  
   \label{ny29}
\end{equation}
subject to the initial condition 
\begin{equation}
   y(x_{0}) = y_{0} > 0, 
   \label{ny30}
\end{equation}
where $C$ is a positive constant. 

\begin{lem} \label{ny:lem3} 
\begin{itemize}
   \item[{\rm (i)}] If $0 < y_{0} < (C/K)$, then there exists a unique 
solution 
$y(x)$ of the initial value problem {\rm (\ref{ny29}), (\ref{ny30})} 
such that $y(x)$ is increasing on $[x_{0},\infty)$ and 
$\lim_{x \to \infty} y(x) = (C/K)$. 
   \item[{\rm (ii)}] If $y_{0} > (C/K)$, then there exists a unique solution 
$y(x)$ of the initial value problem {\rm (\ref{ny29}), (\ref{ny30})} 
such that $y(x)$ is decreasing on $[x_{0},\infty)$ and 
$\lim_{x \to \infty} y(x) = (C/K)$. 
   \item[{\rm (iii)}] If $y_{0} = (C/K)$, then $y(x) \equiv (C/K)$ is 
a solution 
of the initial value problem {\rm (\ref{ny29}), (\ref{ny30})}. 
\end{itemize}
\end{lem}

{\Proof}
Assume that there exists a solution $y(x)$ of the initial value problem 
(\ref{ny29}), (\ref{ny30}) for which $\vert y(x)\vert \neq 0\ (x \geq x_{0})$ 
and $\vert y(x) - (C/K) \vert \neq 0\ (x \geq x_{0})$. 
Then, $y(x) > (C/K)\ (x \geq x_{0})$ or $0 < y(x) < (C/K)\ (x \geq x_{0})$. 
First we suppose that $y(x) > (C/K)\ (x \geq x_{0})$, then we see that 
$y_{0} = y(x_{0}) > (C/K)$ and that 
$$
   y'(x) = -K + \frac{C}{y(x)} = \frac{-Ky(x)+C}{y(x)} < 0, \quad 
   x > x_{0}, 
$$
and therefore $y(x)$ is decreasing on $[x_{0},\infty)$. 
It follows from (\ref{ny29}) that 
$$
    \frac{y(x)}{-Ky(x)+C} y'(x) = 1, \quad x > x_{0}, 
$$
or
\begin{equation}
   \left( 1 + \frac{C}{K}\frac{1}{y(x)- (C/K)}\right)y'(x) = -K, \quad 
   x > x_{0}. 
   \label{ny31}
\end{equation}
Integrating (\ref{ny31}) over $[x_{0},x]$ yields 
$$
   \int_{x_{0}}^{x} \left( 1 + \frac{C}{K}\frac{1}{y(\xi)- (C/K)}\right)
   y'(\xi)d\xi 
   = -K(x - x_{0}), \quad x \geq x_{0}, 
$$
or
$$
   \int_{y(x_{0})}^{y(x)} \left( 1 + \frac{C}{K}\frac{1}{s- (C/K)}\right)ds 
   = -K(x - x_{0}), \quad x \geq x_{0}. 
$$
Hence we obtain 
$$
   \left[ s + \frac{C}{K} \log\left(s - \frac{C}{K}\right)
   \right]_{y_{0}}^{y(x)} = - K(x-x_{0}), \quad x \geq x_{0},   
$$
or equivalently 
\begin{equation}
   y(x) - y_{0} + \frac{C}{K} \log\left(y(x)-\frac{C}{K}\right) 
   - \frac{C}{K} \log\left(y_{0}-\frac{C}{K}\right) = -K(x-x_{0}) 
   \label{ny32}
\end{equation}
for $x \geq x_{0}$. 
Taking the limit as $x \to \infty$ in (\ref{ny32}), we observe that
$$
   \lim_{x \to \infty} \log \left(y(x) - \frac{C}{K}\right) = -\infty
$$
since $y(x) > (C/K)$ and 
$\lim_{x \to \infty} -K(x-x_{0}) = -\infty$. 
Hence we get $\lim_{x \to \infty} y(x) = (C/K)$. 
Next we let $0 < y(x) <(C/K)\ (x \geq x_{0})$, then we deduce that 
$0 < y_{0} = y(x_{0}) < (C/K)$ and that 
$$
   y'(x) = -K + \frac{C}{y(x)} = \frac{-Ky(x)+C}{y(x)} > 0, \quad 
   x > x_{0}, 
$$
and hence $y(x)$ is increasing on $[x_{0}, \infty)$. 
From (\ref{ny29}) we see that
$$
   \left( 1 + \frac{C}{K}\frac{-1}{(C/K) - y(x)}\right)y'(x) = -K, 
   \quad x > x_{0}. 
$$
Integrating the above on $[x_{0}, x]$ and then arguing as in the case where 
$y(x) > (C/K)\ (x \geq x_{0})$, we are led to 
\begin{equation}
   y(x) - y_{0} + \frac{C}{K} \log\left(\frac{C}{K} - y(x)\right) 
   - \frac{C}{K} \log\left(\frac{C}{K} - y_{0}\right) = -K(x-x_{0}) 
   \label{ny33}
\end{equation}
for $x \geq x_{0}$. 
We take the limit as $x \to \infty$ in (\ref{ny33}) to obtain 
$$
   \lim_{x \to \infty} \log \left(\frac{C}{K} - y(x)\right) = -\infty
$$
and hence $\lim_{x \to \infty} y(x) = (C/K)$. 

Conversely we find a solution $y(x)$ of the initial value problem 
(\ref{ny29}), (\ref{ny30}) by utilizing (\ref{ny32}) and (\ref{ny33}). 
First, let $y_{0} > (C/K)$ and consider the equation 
$$
   G(x,y) := y + \frac{C}{K} \log\left(y-\frac{C}{K}\right) 
   + Kx - Kx_{0} - y_{0} 
   - \frac{C}{K} \log\left(y_{0}-\frac{C}{K}\right) = 0 
$$
for $x \geq x_{0},\ y > (C/K)$. 
It is easy to see that 
$$
   G_{y}(x,y) = 1 + \frac{C}{K} \frac{1}{y-(C/K)} 
              = \frac{Ky}{Ky-C} > 0 
$$
for $x \geq x_{0}$, where $G_{y}(x,y) = \frac{\partial G}{\partial y}(x,y)$. 
For every $x \geq x_{0}$ we obtain 
$$
   \lim_{y \to (C/K) + 0} G(x,y) = - \infty, \quad 
   G(x,y_{0}) \geq 0. 
$$
The intermediate value theorem (cf. \cite[Theorem 4.23]{r76}) implies that
for every $x \geq x_{0}$ there exists a unique $y$ for which 
$G(x,y) = 0$ and $(C/K) < y \leq y_{0}$.  Therefore there exists the implicit 
function $y(x)$ such that $G(x,y(x)) = 0$ and $(C/K) < y(x) \leq y_{0}$ for $x \geq x_{0}$. 
Since $G_{y}(x,y) > 0$ for every $x > x_{0}$, 
the implicit function theorem (cf. \cite[Theorem 9.28]{r76}) 
means that $y(x) \in C^{1}(x_{0},\infty)$ and that 
\begin{eqnarray*}
   y'(x) & = & - \frac{G_{x}(x,y(x))}{G_{y}(x,y(x))} 
               = - \frac{K}{Ky(x)/(Ky(x)-C)} \\
         & = & - K + \frac{C}{y(x)}, \quad x > x_{0}, 
\end{eqnarray*}
and therefore $y(x)$ satisfies (\ref{ny29}). 
Since $y(x) > (C/K)\ (x > x_{0})$, we arrive at $y'(x) < 0\ (x > x_{0})$ 
and hence $y(x)$ is decreasing on $[x_{0},\infty)$. 
It is easily verified that $G(x_{0},y(x_{0})) = 0$ implies 
$y(x_{0}) = y_{0}\ (> (C/K))$. 
Next, let $0 < y_{0} < (C/K)$ and treat the equation 
$$
   H(x,y) := y + \frac{C}{K} \log\left(\frac{C}{K}-y\right) 
   + Kx - Kx_{0} - y_{0} 
   - \frac{C}{K} \log\left(\frac{C}{K}-y_{0}\right) = 0 
$$
for $x \geq x_{0},\ 0 < y < (C/K)$. 
We easily see that 
$$
   H_{y}(x,y) = 1 + \frac{C}{K} \frac{-1}{(C/K) - y} 
              = \frac{-Ky}{C - Ky} < 0 
$$
for $x \geq x_{0}$. 
For every $x \geq x_{0}$ we get 
$$
   \lim_{y \to (C/K) - 0} H(x,y) = - \infty, \quad 
   H(x,y_{0}) \geq 0. 
$$
From the intermediate value theorem we see that for every $x \geq x_{0}$ 
there exists a unique $y$ such that $H(x,y) = 0$ and $y_{0} \leq y < (C/K)$. 
Hence there exists the implicit function $y(x)$ such that 
$H(x,y(x)) = 0$ and $y_{0} \leq y(x) < (C/K)$ for $x \geq x_{0}$. 
Since $H_{y}(x,y(x)) < 0$ for every $x > x_{0}$, the implicit function 
theorem implies that $y(x) \in C^{1}(x_{0},\infty)$ and 
\begin{eqnarray*}
   y'(x) & = & - \frac{H_{x}(x,y(x))}{H_{y}(x,y(x))} 
               = - \frac{K}{-Ky(x)/(C-Ky(x))} \\
         & = & - K + \frac{C}{y(x)}, \quad x > x_{0}, 
\end{eqnarray*}
and hence $y(x)$ satisfies (\ref{ny29}). 
Since $y_{0} \leq y(x) < (C/K)\ (x > x_{0})$, we find that 
$y'(x) > 0 \ (x > x_{0})$ 
and that $y(x)$ is increasing on $[x_{0},\infty)$. 
It is obvious that $H(x_{0},y(x_{0})) = 0$ implies 
$y(x_{0}) = y_{0}\ (0 < y_{0} < (C/K))$. 
If $y_{0} = (C/K)$, then it is clear that $y(x) \equiv C/K$ 
is a solution of (\ref{ny29}), (\ref{ny30}). 
\qed

\begin{lem} \label{ny:lem4} 
Assume that one of the following hypotheses {\rm (H$_{1}$)}, 
{\rm (H$_{2}$)} holds{\rm :} 
\begin{itemize}
   \item[{\rm (H$_{1}$)}] $f(x) \in C[x_{0},\alpha)$, 
      $f(x) > 0$ in $[x_{0},\alpha)$ and $f(x)$ is decreasing in 
      $[x_{0},\alpha)${\rm ;} 
   \item[{\rm (H$_{2}$)}] $f(x) \in C[x_{0},\alpha)$, 
      $f(x) > 0$ in $[x_{0},\alpha)$ and $\lim_{x \to \alpha-0} f(x) = 0$. 
\end{itemize} 
Then, for any interval $[\alpha_{1},\alpha_{2}] \subset [x_{0},\alpha)$ 
there exists a point $\tilde{x} \in [x_{0},\alpha)$ such that 
$$
   f(x) > f(\tilde{x}) \quad \mbox{on}\ [\alpha_{1},\alpha_{2}]. 
$$
\end{lem}

{\Proof} 
In the case of (H$_{1}$) we choose $\tilde{x} \in (\alpha_{2},\alpha)$, 
then $f(\alpha_{2}) > f(\tilde{x})$ holds since $f(x)$ is decreasing in 
$[x_{0},\alpha)$. It is easily seen that 
$$
   f(x) \geq f(\alpha_{2}) > f(\tilde{x}) \quad \mbox{on}\ 
   [\alpha_{1},\alpha_{2}]. 
$$
In case (H$_{2}$) holds, the fact that 
$f(x) > 0$ in $[x_{0},\alpha)$ and $\lim_{x \to \alpha-0} f(x) = 0$ 
implies that for 
$\varepsilon_{0} := \min_{\alpha_{1}\leq x \leq \alpha_{2}} f(x) > 0$ 
there exists a number $\delta_{0} > 0$ such that 
$0 < f(x) < \varepsilon_{0}$ holds for any $x \in (\alpha-\delta_{0}, \alpha)$. Hence we conclude that 
$$
   f(x) \geq \min_{\alpha_{1}\leq x \leq \alpha_{2}} f(x) = \varepsilon_{0} 
   > f(\tilde{x}) \quad \mbox{on}\ [\alpha_{1},\alpha_{2}] 
$$
if $\tilde{x} \in (\alpha-\delta_{0}, \alpha)$. 
\qed

\begin{lem} \label{ny:lem5} 
Let $w(x)$ be a local solution of the initial value problem 
{\rm (\ref{ny27}), (\ref{ny28})} defined in $[x_{0}, \tilde{\alpha})$, 
where $\tilde{\alpha}$ is chosen so that $x_{0} < \tilde{\alpha} < \alpha$. 
Assume that there is a point $\tilde{x} \in (\tilde{\alpha}, \alpha)$ 
such that 
$f(x) > f(\tilde{x})$ holds on $[x_{0}, \tilde{\alpha}]$, and let 
$z(x)$ be a solution of the initial value problem
\begin{eqnarray}
   & & z' = - K + \frac{f(\tilde{x})}{z}, \quad x > x_{0}, 
       \label{ny34} \\
   & & z(x_{0}) = z_{0}, 
       \label{ny35}
\end{eqnarray} 
where $0 < z_{0} < w_{0}$.  
Then we deduce that 
\begin{equation}
   w(x) \geq z(x), \quad x_{0} \leq x < \tilde{\alpha}. 
   \label{ny36}
\end{equation} 
\end{lem}

{\Proof} 
First we note that the solution $z(x)$ exists on $[x_{0}, \infty)$ 
by Lemma \ref{ny:lem3}. Suppose to the contrary that (\ref{ny36}) does not hold. Then there is a point $x_{1} \in (x_{0},\tilde{\alpha})$ for which 
$w(x_{1}) < z(x_{1})$. Since $w(x_{0}) > z(x_{0})$, there exists a point 
$x^{*} \in (x_{0},x_{1})$ such that $w(x^{*}) = z(x^{*})$ and 
$w(x) < z(x)$ for $x^{*} < x \leq x_{1}$. 
Since $z(x^{*}) - w(x^{*}) = 0$ and $z(x) - w(x) > 0$ for 
$x^{*} < x \leq x_{1}$, we see that $z'(x^{*}) - w'(x^{*}) \geq 0$, i.e., 
$z'(x^{*}) \geq w'(x^{*})$. 
On the other hand, we obtain
\begin{eqnarray*}
   z'(x^{*}) - w'(x^{*}) 
   & = & \left(-K + \frac{f(\tilde{x})}{z(x^{*})}\right) 
         - \left(-K + \frac{f(x^{*})}{w(x^{*})}\right) \\
   & = & \frac{f(\tilde{x})}{z(x^{*})} 
         - \frac{f(x^{*})}{w(x^{*})} 
         = \frac{f(\tilde{x}) - f(x^{*})}{w(x^{*})} < 0
\end{eqnarray*}
since $f(x^{*}) - f(\tilde{x}) > 0$ by the hypothesis. 
This is a contradiction, and the proof is complete. 
\qed

\begin{thm} \label{ny:thm3} 
Assume that either {\rm (H$_{1}$)} or {\rm (H$_{2}$)} 
in Lemma {\rm \ref{ny:lem4}} 
holds. Then there exists a unique global {\rm ({\it positive})} 
solution $w(x)$ of the 
initial value problem {\rm (\ref{ny27}), (\ref{ny28})} 
{\rm (or (\ref{ny26}), (\ref{ny28}))}. 
\end{thm}

{\Proof}
It follows from Theorem \ref{ny:thm2} that there exists a unique local 
(positive) solution $w(x)$ of the problem (\ref{ny27}), (\ref{ny28}). 
Suppose to the contrary that the solution $w(x)$ has the maximal interval of 
existence $[x_{0},\alpha^{*})$, where $x_{0} < \alpha^{*} < \alpha$. 
Lemma \ref{ny:lem4} implies that for the interval $[x_{0},\alpha^{*}]$ 
there is a point $\tilde{x} \in (\alpha^{*}, \alpha)$ such that 
$f(x) > f(\tilde{x})$ on $[x_{0},\alpha^{*}]$. 
From Lemma \ref{ny:lem5} we see that 
\begin{equation}
   w(x) \geq z(x), \quad x_{0} \leq x < \alpha^{*}, 
   \label{ny37}
\end{equation}
where $z(x)$ is a positive solution of the initial value problem 
(\ref{ny34}), (\ref{ny35}). 
Since (\ref{ny27}) is equivalent to (\ref{ny26}) in $[x_{0}, \alpha^{*})$, 
we get 
$$
   \frac{1}{2}\bigl(w(x)^{2}\bigr)' + K w(x) = f(x), \quad 
   x_{0} \leq x < \alpha^{*}, 
$$
or
$$
   \bigl(w(x)^{2}\bigr)' = -2 K w(x) + 2f(x) \leq 2f(x), \quad 
   x_{0} \leq x < \alpha^{*}. 
$$
Integrating the above inequality on $[x_{0}, x]$ yields 
\begin{equation}
    w(x) \leq \left(w_{0}^{2} + 2\int_{x_{0}}^{x} f(s)ds\right)^{1/2}, 
   \quad x_{0} \leq x < \alpha^{*}. 
   \label{ny38}
\end{equation}
It can be shown from (\ref{ny37}) and (\ref{ny38}) that 
$$
   z(x) \leq w(x) \leq 
   \left(w_{0}^{2} + 2\int_{x_{0}}^{x} f(s)ds\right)^{1/2}, \quad 
   x_{0} \leq x < \alpha^{*}. 
$$
Let $\{\alpha_{n}\}_{n=1}^{\infty}$ be a sequence such that 
$\alpha_{n} < \alpha^{*}$ and $\lim_{n\to \infty} \alpha_{n} = \alpha^{*}$. 
Since $z(x)$ is a positive continuous function in 
$[x_{0},\infty)$ and 
$$
   \left(w_{0}^{2} + 2\int_{x_{0}}^{x} f(s)ds\right)^{1/2} 
   \leq 
   \left(w_{0}^{2} + 2\int_{x_{0}}^{\alpha^{*}} f(s)ds\right)^{1/2}, 
$$
we see that the sequence $\{w(\alpha_{n})\}_{n=1}^{\infty}$ is bounded. 
Hence, $\{w(\alpha_{n})\}_{n=1}^{\infty}$ has a convergent 
subsequence $\{w(\alpha_{n(k)})\}_{k=1}^{\infty}$, and let 
$\lim_{k \to \infty} w(\alpha_{n(k)}) = \ell$. 
Then equation (\ref{ny27}) has a solution $\tilde{w}(x)$ passing through 
$(\alpha^{*},\ell)$ which exists on some interval $[\alpha^{*},\alpha^{*}+\beta]$ 
\ $(\beta > 0)$, and hence 
the solution $w(x)$ may be continued to the right of 
$\alpha^{*}$ (see, e.g., Coddington and Levinson 
\cite[Theorem 4.1]{cl55}). 
This contradicts the fact that $[x_{0},\alpha^{*})$ is the maximal 
interval of existence, and the proof is complete. 
\qed


\section{Exact solution of initial value problem for SEIR 
differential system (II)} 

In this section we give the exact solution of the initial value 
problem (\ref{ny1})--(\ref{ny5}) based on the results of Section 3, 
and show that the parametric solution obtained in Section 2 
can be derived by solving some linear differential system including 
a positive solution of an Abel differential equation. 

\begin{lem} \label{ny:lem6} Under 
the hypothesis {\rm (A$_{4}$)}, 
the transcendental equation 
\begin{equation}
   x = N - \tilde{S}e^{(\beta/\gamma)\tilde{R}}
                 e^{-(\beta/\gamma)x} 
   \label{ny39}
\end{equation}
has a unique solution $x = \alpha$ such that
$$
   \tilde{R} < \alpha < N 
$$
{\rm ({\it cf}. Figure 1)}. 
\end{lem}

\begin{figure}[ht]
  \begin{center}
    \includegraphics[height=4cm]{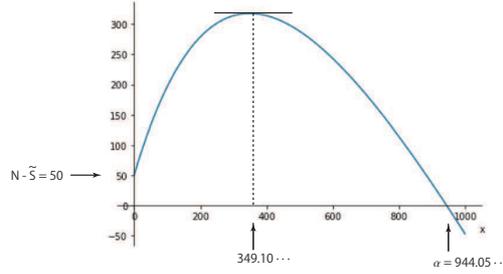}
    \caption{Variation of $N-\tilde{S}e^{-(\beta/\gamma)x}-x$ 
    for $N=1000, \tilde{S}=950, \tilde{R}=0, \beta=0.3/1000$ and $\gamma=0.1$. 
    In this case we find that $N-\tilde{S}=50$, 
    $0 < \alpha = 944.05\cdots < 1000$ and $\psi'(\xi)=0$ for 
    $\xi = 349.10\cdots$.  }
  \end{center}
\end{figure}

{\Proof}
The hypothesis (A$_{4}$) implies that 
$$
   N - \tilde{S}e^{(\beta/\gamma)\tilde{R}} > 0. 
$$
We define the sequence $\{a_{n}\}_{n=1}^{\infty}$ by
\begin{eqnarray}
   a_{1}   & = & \tilde{a}\ 
                 \left(0 < \tilde{a} \leq N 
                 - \tilde{S}e^{(\beta/\gamma)\tilde{R}}\right), \nonumber\\
   a_{n+1} & = & N - \tilde{S}e^{(\beta/\gamma)\tilde{R}} 
                 e^{-(\beta/\gamma)a_{n}}\ (n=1,2,...). \label{ny40}
\end{eqnarray}
It is easy to see that 
$$
   a_{1} = \tilde{a} \leq N - \tilde{S}e^{(\beta/\gamma)\tilde{R}}
   \leq N - \tilde{S}e^{(\beta/\gamma)\tilde{R}}e^{-\frac{\beta}{\gamma}a_{1}}
   = a_{2}. 
$$
If $a_{n+1} \geq a_{n}$, then 
\begin{eqnarray*}
   a_{n+2} - a_{n+1} 
   & = & N - \tilde{S}e^{(\beta/\gamma)\tilde{R}}
         e^{-(\beta/\gamma)a_{n+1}} 
         - \left(N - \tilde{S}e^{(\beta/\gamma)\tilde{R}}
         e^{-(\beta/\gamma)a_{n}}\right) \\
   & = & \tilde{S}e^{(\beta/\gamma)\tilde{R}}
         \left(e^{-(\beta/\gamma)a_{n}}-e^{-(\beta/\gamma)a_{n+1}}
         \right) \\
   & \geq & 0. 
\end{eqnarray*}
Therefore we obtain $a_{n+2} \geq a_{n+1}$, and hence the sequence $\{a_{n}\}$ 
is nondecreasing by the mathematical induction. 
We easily see that the sequence $\{a_{n}\}$ 
is bounded since
$$
   \vert a_{n+1} \vert \leq N + \tilde{S}e^{(\beta/\gamma)\tilde{R}}
                          e^{-(\beta/\gamma)a_{n}} 
   \leq N + \tilde{S}e^{(\beta/\gamma)\tilde{R}}. 
$$
Consequently there exists $\lim_{n \to \infty} a_{n} = \alpha$. 
Taking the limit as $n \to \infty$ in (\ref{ny40}), we get 
$$
   \alpha = N - \tilde{S}e^{(\beta/\gamma)\tilde{R}}
                 e^{-(\beta/\gamma)\alpha}. 
$$
The uniqueness of $\alpha$ follows the fact that 
the straight line $y = N-x$ and the exponential curve 
$y = \tilde{S}e^{(\beta/\gamma)\tilde{R}}e^{-(\beta/\gamma)x}$ 
has only one intersecting point in $0 < x < N$ in view of 
the inequality $N > \tilde{S}e^{(\beta/\gamma)\tilde{R}}$, 
and hence the solution $\alpha$ of 
the transcendental equation (\ref{ny39}) is unique. 
The inequality $\tilde{R} < \alpha < N$ follows from the inequalities 
\begin{eqnarray*}
   \alpha & = & N - \tilde{S}e^{(\beta/\gamma)\tilde{R}}
                e^{-(\beta/\gamma)\alpha} < N, \\
   \alpha & = & N - \tilde{S}e^{(\beta/\gamma)\tilde{R}}
                e^{-(\beta/\gamma)\alpha}
                > N - \tilde{S}e^{(\beta/\gamma)\tilde{R}} > \tilde{R}.
\end{eqnarray*}
\qed

We assume that the following hypothesis 
\begin{itemize}
   \item[{\rm (A$_{5}$)}] $\displaystyle \tilde{S} < 
                     \frac{\gamma}{\beta}
                     e^{(\beta/\gamma)(\alpha-\tilde{R})}$ 
\end{itemize}
holds in the rest of this paper. 
We note that (A$_{5}$) is equivalent to the following 
\begin{itemize}
   \item[(A$_{5}'$)] $\displaystyle \frac{\gamma}{\beta} > N - \alpha$ 
\end{itemize}
in light of $\tilde{S}e^{(\beta/\gamma)\tilde{R}}e^{-(\beta/\gamma)\alpha} 
= N - \alpha$. 

\begin{rem} \label{ny:rem2} \rm 
Combining the hypotheses (A$_{2}$) and (A$_{3}$), we obtain 
$$
   \tilde{S} > \frac{\delta \tilde{E}}{\beta \tilde{I}} 
             > \frac{\gamma}{\beta}. 
$$
\end{rem}

\begin{lem} \label{ny:lem7} 
There exists a unique positive solution $w(x)$ of 
the initial value problem for the Abel differential equation 
\begin{equation}
   \hspace*{0ex} w'w + \frac{\beta(\gamma+\delta)}{\gamma}w 
       = \frac{\beta \delta}{\gamma}
         \left(\beta N - \beta \tilde{S}e^{(\beta/\gamma)\tilde{R}}
               e^{-(\beta/\gamma)x} - \beta x \right), 
         x \in (\tilde{R},\alpha), 
         \label{ny41} 
\end{equation}
subject to the initial condition
\begin{equation}
   w(\tilde{R}) = \beta \tilde{I}. 
       \label{ny42} 
\end{equation}
\end{lem}

\begin{figure}[ht]
  \begin{center}
    \includegraphics[height=5cm]{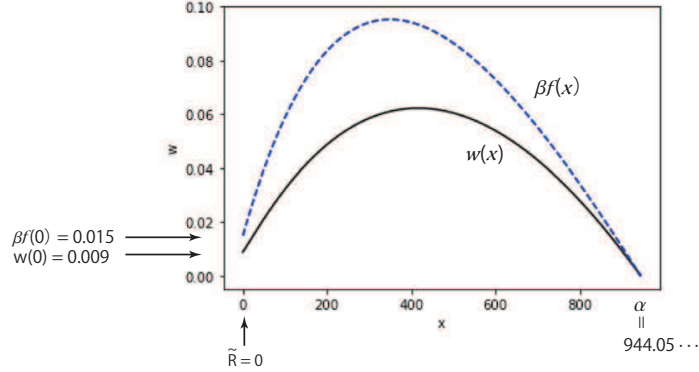}
    \caption{Variations of $\beta f(x)$ (dashed curve), and $w(x)$ (solid curve) obtained by the numerical integration of the initial value problem (\ref{ny41}), (\ref{ny42}), for $N=1000, \tilde{S}=950, \tilde{E}=20, \tilde{I}=30, \tilde{R}=0, \beta=0.3/1000, \gamma=0.1, \delta=0.2$ and $\alpha=944.05\cdots$. 
In this case we have $\beta f(0) = \beta N - \beta\tilde{S} = 0.015$ and 
$w(0) = \beta \tilde{I} = 0.009$. }
  \end{center}
\end{figure}

{\Proof}
Let 
$$
  f(x) := N - \tilde{S}e^{(\beta/\gamma)\tilde{R}}
                 e^{-(\beta/\gamma)x} - x. 
$$
Since $f'(x) = 0$ for 
$$
   x = \tilde{x} = \frac{\gamma}{\beta} 
       \log\ \bigl((\beta/\gamma)\tilde{S}e^{(\beta/\gamma)\tilde{R}}\bigr), 
$$
we see that $\tilde{R} < \tilde{x} < \alpha$ by means of 
(A$_{5}$) and Remark \ref{ny:rem2}, and that 
$f'(x) > 0$ for $\tilde{R} < x < \tilde{x}$ and $f'(x) < 0$ 
for $\tilde{x} < x < \alpha$. Hence, $f(x)$ is increasing in 
$[\tilde{R}, \tilde{x})$ and decreasing in $(\tilde{x},\alpha)$. 
Since $f(\tilde{R}) = N-\tilde{S}-\tilde{R} = \tilde{E}+\tilde{I} > 0$ and 
$\lim_{x \to \alpha-0} f(x) = 0$ by Lemma \ref{ny:lem6}, 
it follows that $f(x) \in C[\tilde{R},\alpha)$, $f(x) > 0$ in 
$[\tilde{R},\alpha)$ and $\lim_{x \to \alpha-0} f(x) = 0$. 
Therefore there exists a unique positive solution $w(x)$ 
of the initial value problem (\ref{ny41}), (\ref{ny42}) 
by Theorem \ref{ny:thm3} in Section 3 
({\it cf}. Figure 2). 

\qed

\begin{lem}  \label{ny:lem8} 
There exists a unique solution $\psi(u)$ of the initial value problem 
\begin{eqnarray}
   & & \hspace*{-7ex} \psi'\psi - \frac{\gamma+\delta}{u}\psi 
       = - \delta\,\frac{\beta N - \beta \tilde{S}e^{(\beta/\gamma)\tilde{R}}u
       + \gamma \log u}{u}, 
       u \in (e^{-\frac{\beta}{\gamma}\alpha}, 
              e^{-\frac{\beta}{\gamma}\tilde{R}}), 
       \label{ny43} \\
   & & \hspace*{-7ex} \psi\bigl(e^{-(\beta/\gamma)\tilde{R}}\bigr) 
       = \beta \tilde{I} 
       \label{ny44}
\end{eqnarray}
satisfying
\begin{equation}
   \psi(u) > 0\ {\rm in}\ 
       (e^{-(\beta/\gamma)\alpha}, e^{-(\beta/\gamma)\tilde{R}}] 
       \label{ny45}
\end{equation}
{\rm ({\it cf}. Figure 3)}. 
\end{lem}

\begin{figure}[ht]
  \begin{center}
    \includegraphics[height=5cm]{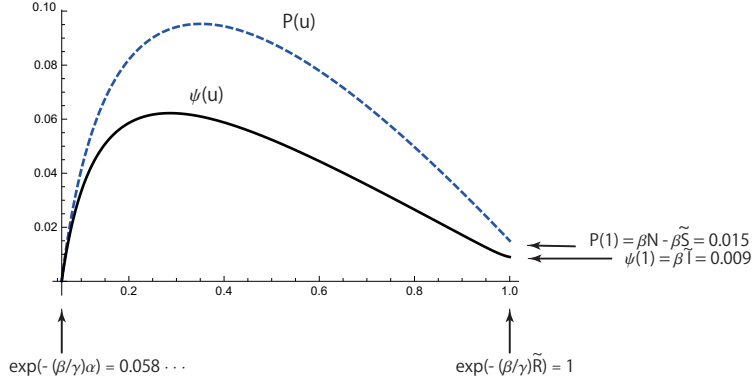}
    \caption{Variations of $P(u) := \beta N-\beta \tilde{S}
    e^{(\beta/\gamma)\tilde{R}}u+\gamma \log u$ (dashed curve) and 
    $\psi(u)$ (solid curve) obtained by the numerical integration of the initial    value problem (\ref{ny43}), (\ref{ny44}) for $N=1000, \tilde{S}=950, 
    \tilde{E}=20, 
    \tilde{I}=30, \tilde{R}=0, \beta=0.3/1000, \gamma=0.1, \delta=0.2$ and 
    $\alpha=944.05\cdots$. 
    In this case we have $e^{-(\beta/\gamma)\alpha}=0.058\cdots$, 
    $e^{-(\beta/\gamma)\tilde{R}}=1$, 
    $P(1)=\beta N-\beta\tilde{S}=0.015$ and 
    $\psi(1)=\beta \tilde{I}=0.009$.  }
  \end{center}
\end{figure}

{\Proof}
Let $w(x)$ be a unique positive solution of the initial value problem 
(\ref{ny41}), (\ref{ny42}). 
We define $\psi(u)$ by
$$
   \psi(u) := w\left(-\frac{\gamma}{\beta}\log u\right) 
$$
and find that 
$$
   \psi'(u) = w'\left(-\frac{\gamma}{\beta}\log u\right)
              \left(- \frac{\gamma}{\beta}\frac{1}{u}\right), 
$$
and hence 
\begin{eqnarray*}
   &   & \psi'(u)\psi(u) \\
   & = & - \frac{\gamma}{\beta}\,\frac{1}{u}\,
         w'\left(-\frac{\gamma}{\beta}\log u\right)
         w\left(-\frac{\gamma}{\beta}\log u\right) \\
   & = & - \frac{\gamma}{\beta}\,\frac{1}{u} 
         \left[- \frac{\beta(\gamma+\delta)}{\gamma}
               w\left(-\frac{\gamma}{\beta}\log u\right) \right. \\
   &   & \hspace*{15ex} \left.  + \frac{\beta\delta}{\gamma}
           \left(\beta N - \beta\tilde{S}e^{(\beta/\gamma)\tilde{R}}u 
           + \gamma \log u\right)\right] \\
   & = & \frac{\gamma+\delta}{u}\psi(u) 
         -\delta\,\frac{\beta N - \beta\tilde{S}e^{(\beta/\gamma)\tilde{R}}u 
         + \gamma \log u}{u}  \\
   &   & \hspace*{30ex} \mbox{for}\ 
         e^{-(\beta/\gamma)\alpha} < u < e^{-(\beta/\gamma)\tilde{R}} 
\end{eqnarray*}
by means of (\ref{ny41}). Hence $\psi(u)$ satisfies 
the Abel differential equation (\ref{ny43}). 
It is easily seen from (\ref{ny42}) that 
$$
   \psi\bigl(e^{-(\beta/\gamma)\tilde{R}}\bigr) 
   = w(\tilde{R}) = \beta \tilde{I} 
$$
and therefore (\ref{ny44}) is satisfied. 
The uniqueness of $\psi(u)$ follows from that of $w(x)$. 
It is clear that 
$$
   \psi(u) > 0\ \mbox{in}\ 
   (e^{-(\beta/\gamma)\alpha}, e^{-(\beta/\gamma)\tilde{R}}]
$$
since $\psi(u) = w\bigl(-(\gamma/\beta)\log u\bigr)$ and 
$w(x) > 0$ in $[\tilde{R},\alpha)$, and hence 
(\ref{ny45}) is satisfied. 
\qed

\begin{lem}  \label{ny:lem9} 
The unique positive solution $\psi(u)$ of the initial value problem 
{\rm (\ref{ny43}), (\ref{ny44})} satisfies 
the following relation 
\begin{eqnarray}
   \psi(u) 
   & = & \beta N-\beta\tilde{S}e^{(\beta/\gamma)\tilde{R}}u+\gamma \log u 
         \nonumber \\
   &   & -\beta \left( \tilde{E}e^{-\delta\varphi(u)} 
         + \tilde{S}e^{(\beta/\gamma)\tilde{R}}e^{-\delta\varphi(u)}
           \int_{u}^{e^{-(\beta/\gamma)\tilde{R}}}
           e^{\delta\varphi(v)}dv \right) 
         \label{ny46} 
\end{eqnarray}
for $e^{-(\beta/\gamma)\alpha} < u \leq e^{-(\beta/\gamma)\tilde{R}}$, 
where 
\begin{equation}
   \varphi(u) := \int_{u}^{e^{-(\beta/\gamma)\tilde{R}}} 
                     \frac{d\xi}{\xi\psi(\xi)}. 
                     \label{ny47}
\end{equation}
Conversely, the function $\psi(u)$ satisfying 
{\rm (\ref{ny45}), (\ref{ny46})} 
is a  solution of the initial value problem 
{\rm (\ref{ny43}), (\ref{ny44})}. 
\end{lem}

{\Proof}
First we note that 
(\ref{ny46}) is some kind of integral equation of 
$\psi(u)$, in view of (\ref{ny47}). 
Let $\psi(u)$ be the unique positive solution of the problem 
(\ref{ny43}), (\ref{ny44}), and define 
$z(u)$ by
\begin{equation}
   z(u) := \psi(u) - \bigl( 
           \beta N-\beta\tilde{S}e^{(\beta/\gamma)\tilde{R}}u+\gamma \log u 
           \bigr). 
           \label{ny48}
\end{equation}
Dividing (\ref{ny43}) by $\psi(u)$ yields 
\begin{eqnarray}
   \psi'(u) 
   & = & \frac{\gamma+\delta}{u} 
         - \delta\,\frac{\beta N - \beta \tilde{S}e^{(\beta/\gamma)\tilde{R}}u
         + \gamma \log u}{u\psi(u)} 
         \nonumber \\
   & = & \frac{\gamma}{u} 
         - \delta\,\frac{\beta N - \beta \tilde{S}e^{(\beta/\gamma)\tilde{R}}u
         + \gamma \log u - \psi(u)}{u\psi(u)} 
         \nonumber \\
   & = & \frac{\gamma}{u} 
         + \delta\,\frac{z(u)}{u\psi(u)}. 
         \label{ny49}
\end{eqnarray}
On the other hand, we differentiate (\ref{ny48}) to obtain 
\begin{equation}
   \psi'(u) = - \beta \tilde{S}e^{(\beta/\gamma)\tilde{R}} 
              + \frac{\gamma}{u} + z'(u). 
              \label{ny50}
\end{equation}
Combining (\ref{ny49}) with (\ref{ny50}) yields 
$$
   z'(u) - \frac{\delta}{u\psi(u)}z(u) 
   = \beta \tilde{S}e^{(\beta/\gamma)\tilde{R}} 
$$
or
\begin{equation}
   z'(u) + \delta \varphi'(u)z(u) 
   = \beta \tilde{S}e^{(\beta/\gamma)\tilde{R}} 
   \label{ny51}
\end{equation}
which is a linear differential equation of first order. 
It is obvious that 
\begin{eqnarray}
   z\bigl(e^{-(\beta/\gamma)\tilde{R}}\big) 
   & = & \psi\bigl(e^{-(\beta/\gamma)\tilde{R}}\big) 
         - \beta\bigl(N - \tilde{S} - \tilde{R}\bigr) 
         \nonumber \\
   & = & \beta \tilde{I} - \beta\bigl(N - \tilde{S} - \tilde{R}\bigr) 
         \nonumber \\
   & = & - \beta \tilde{E}. 
         \label{ny52}
\end{eqnarray}
Now we solve the initial value problem 
(\ref{ny51}), (\ref{ny52}). 
Multiplying (\ref{ny51}) by $e^{\delta\varphi(u)}$ gives 
$$
   \bigl(e^{\delta\varphi(u)}z(u)\bigr)' 
   = \beta \tilde{S}e^{(\beta/\gamma)\tilde{R}}e^{\delta\varphi(u)} 
$$
and then integrating the above on $[u, e^{-(\beta/\gamma)\tilde{R}}]$ yields 
$$
   z\bigl(e^{-(\beta/\gamma)\tilde{R}}\bigr) 
   - e^{\delta\varphi(u)}z(u) 
   = \beta \tilde{S}e^{(\beta/\gamma)\tilde{R}}
     \int_{u}^{e^{-(\beta/\gamma)\tilde{R}}}e^{\delta\varphi(v)}dv. 
$$
Taking account of (\ref{ny52}), we obtain 
\begin{equation}
   z(u) = -\beta \left( \tilde{E}e^{-\delta\varphi(u)} 
         + \tilde{S}e^{(\beta/\gamma)\tilde{R}}e^{-\delta\varphi(u)}
           \int_{u}^{e^{-(\beta/\gamma)\tilde{R}}}
           e^{\delta\varphi(v)}dv \right). 
        \label{ny53}
\end{equation}
Combining (\ref{ny48}) with (\ref{ny53}), we conclude that 
$\psi(u)$ satisfies (\ref{ny46}) for 
$e^{-(\beta/\gamma)\alpha} < u < e^{-(\beta/\gamma)\tilde{R}}$. 
If $u = e^{-(\beta/\gamma)\tilde{R}}$, then 
$\psi\bigl(e^{-(\beta/\gamma)\tilde{R}}\bigr) = \beta \tilde{I}$ 
by (\ref{ny44}) and the right hand side of (\ref{ny46}) with 
$u = e^{-(\beta/\gamma)\tilde{R}}$ is equal to 
$\beta N - \beta \tilde{S} - \beta \tilde{R} - \beta \tilde{E} 
= \beta \tilde{I}$. Therefore (\ref{ny46}) holds for 
$u = e^{-(\beta/\gamma)\tilde{R}}$. 

Conversely we suppose that the function $\psi(u)$ satisfies (\ref{ny45}), 
(\ref{ny46}), and let 
$e^{-(\beta/\gamma)\alpha} < u < e^{-(\beta/\gamma)\tilde{R}}$. 
Differentiating (\ref{ny46}) with respect to $u$ yields 
\begin{eqnarray}
   \psi'(u) 
   & = & - \beta \tilde{S}e^{(\beta/\gamma)\tilde{R}} 
         + \frac{\gamma}{u} 
         - \beta \tilde{E}e^{-\delta\varphi(u)}\bigl(-\delta \varphi'(u)\bigr) 
         \nonumber \\  
   &   & \quad - \beta \tilde{S}e^{(\beta/\gamma)\tilde{R}}
         \left( e^{-\delta\varphi(u)}\bigl(-\delta \varphi'(u)\bigr)
         \int_{u}^{e^{-(\beta/\gamma)\tilde{R}}}
           e^{\delta\varphi(v)}dv - 1 \right) \nonumber \\
   & = & \frac{\gamma}{u} - \beta \delta \tilde{E}e^{-\delta\varphi(u)}
         \frac{1}{u\psi(u)} \nonumber \\
   &   & \quad 
         - \beta\delta \tilde{S}e^{(\beta/\gamma)\tilde{R}}
         \frac{1}{u\psi(u)} e^{-\delta\varphi(u)} 
         \int_{u}^{e^{-(\beta/\gamma)\tilde{R}}}
           e^{\delta\varphi(v)}dv. 
         \label{ny54}
\end{eqnarray}
It is easily seen from (\ref{ny46}) that 
\begin{eqnarray}
   &   & - \beta \tilde{S}e^{(\beta/\gamma)\tilde{R}}e^{-\delta\varphi(u)}
           \int_{u}^{e^{-(\beta/\gamma)\tilde{R}}}
           e^{\delta\varphi(v)}dv  \nonumber \\
   & = & \psi(u) 
         - \bigl(\beta N-\beta\tilde{S}e^{(\beta/\gamma)\tilde{R}}u 
         +\gamma \log u \bigr) 
         + \beta \tilde{E}e^{-\delta\varphi(u)}. 
         \label{ny55} 
\end{eqnarray} 
We combine (\ref{ny54}) with (\ref{ny55}) to obtain 
\begin{eqnarray*}
   \psi'(u) 
  & = & \frac{\gamma}{u} - \beta \delta \tilde{E}e^{-\delta\varphi(u)}
         \frac{1}{u\psi(u)}  \\
  &   & \quad 
        + \delta\,\frac{\psi(u) 
        - \bigl(\beta N-\beta\tilde{S}e^{(\beta/\gamma)\tilde{R}}u 
        +\gamma \log u \bigr) 
        + \beta \tilde{E}e^{-\delta\varphi(u)}}{u\psi(u)}  \\
  & = & \frac{\gamma}{u} 
        - \delta\,\frac{
        \beta N-\beta\tilde{S}e^{(\beta/\gamma)\tilde{R}}u 
        +\gamma \log u  - \psi(u)} 
        {u\psi(u)}  \\
  & = & \frac{\gamma+\delta}{u} 
        - \delta\,\frac{
        \beta N-\beta\tilde{S}e^{(\beta/\gamma)\tilde{R}}u 
        +\gamma \log u }
        {u\psi(u)} 
\end{eqnarray*}
and consequently, $\psi(u)$ satisfies (\ref{ny43}) 
for $e^{-(\beta/\gamma)\alpha} < u < e^{-(\beta/\gamma)\tilde{R}}$. 
It can be shown from (\ref{ny46}) that 
$$
   \psi\bigl(e^{-(\beta/\gamma)\tilde{R}}\bigr) 
   = \beta N - \beta \tilde{S} - \beta \tilde{R} - \beta \tilde{E} 
   = \beta \tilde{I} 
$$
in light of $\varphi\bigl(e^{-(\beta/\gamma)\tilde{R}}\bigr) = 0$, 
and hence (\ref{ny44}) is satisfied. 
\qed

\begin{rem} \label{ny:rem3} \rm 
The relation (\ref{ny46}) in Lemma \ref{ny:lem9} can be derived 
from (\ref{ny43}) by another technique. 
We find from (\ref{ny43}) that 
$$
   \psi'(u)\psi(u) - \frac{\gamma}{u}\,\psi(u) 
   = - \delta\,\frac{\beta N - \beta \tilde{S}e^{(\beta/\gamma)\tilde{R}}u
       - (\psi(u)-\gamma \log u)}{u}. 
$$
Letting $g(u) := \psi(u) - \gamma \log u$, we obtain 
$$
   g'(u)\psi(u) = - \delta\,\frac{\beta N 
                  - \beta \tilde{S}e^{(\beta/\gamma)\tilde{R}}u - g(u)}{u}, 
$$
or 
$$
   g'(u) - \frac{\delta}{u\psi(u)} g(u) 
   = - \delta\,\frac{\beta N 
                  - \beta \tilde{S}e^{(\beta/\gamma)\tilde{R}}u}{u\psi(u)} 
$$
which is equal to 
\begin{eqnarray*}
   g'(u) + \delta \varphi'(u)g(u) 
   & = & - \delta\,\frac{\beta N 
                  - \beta \tilde{S}e^{(\beta/\gamma)\tilde{R}}u}{u\psi(u)} \\
   & = & \delta \varphi'(u)\bigl(\beta N 
                  - \beta \tilde{S}e^{(\beta/\gamma)\tilde{R}}u \bigr). 
\end{eqnarray*}
Multiplying the above linear differential equation of first order 
by $e^{\delta \varphi(u)}$ and then integrating over 
$[u,e^{-(\beta/\gamma)\tilde{R}}]$, we get 
\begin{eqnarray*}
   g(u) 
   & = & \beta\bigl(\tilde{I}+\tilde{R}\bigr)e^{-\delta\varphi(u)} \\
   &   & \quad - e^{-\delta\varphi(u)}
         \int_{u}^{e^{-(\beta/\gamma)\tilde{R}}} 
         \bigl(\beta N - \beta \tilde{S}e^{(\beta/\gamma)\tilde{R}}v\bigr) 
         \bigl(e^{\delta\varphi(v)}\bigr)'dv. 
\end{eqnarray*}
Since 
\begin{eqnarray*}
   &   & \int_{u}^{e^{-(\beta/\gamma)\tilde{R}}} 
         \bigl(\beta N - \beta \tilde{S}e^{(\beta/\gamma)\tilde{R}}v\bigr) 
         \bigl(e^{\delta\varphi(v)}\bigr)'dv \\
   & = & \beta (N - \tilde{S}) 
         - \bigl(\beta N - \beta \tilde{S}e^{(\beta/\gamma)\tilde{R}}u\bigr)
         e^{\delta\varphi(u)} 
         + \beta \tilde{S}e^{(\beta/\gamma)\tilde{R}}
           \int_{u}^{e^{-(\beta/\gamma)\tilde{R}}}e^{\delta\varphi(v)}dv 
\end{eqnarray*}
by integrating by parts, we arrive at
\begin{eqnarray*}
   g(u) 
  & = & \beta N - \beta \tilde{S}e^{(\beta/\gamma)\tilde{R}}u 
        - \beta \tilde{E}e^{-\delta \varphi(u)} \\
  &   & \quad - \beta \tilde{S}e^{(\beta/\gamma)\tilde{R}} 
              e^{-\delta \varphi(u)} \int_{u}^{e^{-(\beta/\gamma)\tilde{R}}} 
              e^{\delta \varphi(v)}dv 
\end{eqnarray*}
which is equivalent to (\ref{ny46}) in light of 
$g(u) = \psi(u) - \gamma \log u$. 
\end{rem}

\begin{prop} \label{ny:prop1} 
Let $\psi(u)$ be the unique positive solution of the initial value 
problem {\rm (\ref{ny43}), (\ref{ny44})}, then 
the following inequalities hold{\rm :}
\begin{eqnarray}
   &   & \beta N-\beta\tilde{S}e^{(\beta/\gamma)\tilde{R}}u+\gamma \log u 
         > \psi(u) > 0, 
         \label{ny56} \\
   &   & \beta N-\beta\tilde{S}e^{(\beta/\gamma)\tilde{R}}u+\gamma \log u 
         \nonumber \\
   &   & > \beta \left( \tilde{E}e^{-\delta\varphi(u)} 
         + \tilde{S}e^{(\beta/\gamma)\tilde{R}}e^{-\delta\varphi(u)}
           \int_{u}^{e^{-(\beta/\gamma)\tilde{R}}}
           e^{\delta\varphi(v)}dv \right) > 0 
         \label{ny57}
\end{eqnarray}
for $e^{-(\beta/\gamma)\alpha} < u \leq e^{-(\beta/\gamma)\tilde{R}}$. 
\end{prop}

{\Proof}
Since $\psi(u) > 0$ in 
$(e^{-(\beta/\gamma)\alpha}, e^{-(\beta/\gamma)\tilde{R}}]$, 
the relation (\ref{ny46}) in Lemma \ref{ny:lem9} implies
\begin{eqnarray*}
   &   & \beta N-\beta\tilde{S}e^{(\beta/\gamma)\tilde{R}}u+\gamma \log u \\
   & > & \beta \left( \tilde{E}e^{-\delta\varphi(u)} 
         + \tilde{S}e^{(\beta/\gamma)\tilde{R}}e^{-\delta\varphi(u)}
           \int_{u}^{e^{-(\beta/\gamma)\tilde{R}}}
           e^{\delta\varphi(v)}dv \right) 
\end{eqnarray*}
for $e^{-(\beta/\gamma)\alpha} < u \leq e^{-(\beta/\gamma)\tilde{R}}$. 
It is obvious that
\begin{equation}
   \tilde{E}e^{-\delta\varphi(u)} 
         + \tilde{S}e^{(\beta/\gamma)\tilde{R}}e^{-\delta\varphi(u)}
           \int_{u}^{e^{-(\beta/\gamma)\tilde{R}}}
           e^{\delta\varphi(v)}dv
   > 0  
   \label{ny58}
\end{equation}
for $e^{-(\beta/\gamma)\alpha} < u \leq e^{-(\beta/\gamma)\tilde{R}}$, 
and hence (\ref{ny57}) holds. 
Since (\ref{ny58}) holds, the relation (\ref{ny46}) means that 
$$
   \beta N-\beta\tilde{S}e^{(\beta/\gamma)\tilde{R}}u+\gamma \log u 
   > \psi(u) > 0 \quad \mbox{for}\ 
   e^{-(\beta/\gamma)\alpha} < u \leq e^{-(\beta/\gamma)\tilde{R}} 
$$
which is the desired inequality (\ref{ny56}). 
\qed

\begin{prop} \label{ny:prop2} 
Let $\psi(u)$ be the unique positive solution of the initial value 
problem {\rm (\ref{ny43}), (\ref{ny44})}, then we have 
\begin{eqnarray}
   &   & \hspace*{-5ex} \lim_{u \to e^{-(\beta/\gamma)\alpha}+0} \psi(u) = 0, 
         \label{ny59} \\
   &   & \hspace*{-5ex} 
         \lim_{u \to e^{-(\beta/\gamma)\alpha}+0} 
         \left( \tilde{E}e^{-\delta\varphi(u)} 
         + \tilde{S}e^{(\beta/\gamma)\tilde{R}}e^{-\delta\varphi(u)}
           \int_{u}^{e^{-(\beta/\gamma)\tilde{R}}}
           e^{\delta\varphi(v)}dv\right)  = 0 
         \label{ny60} 
\end{eqnarray}
{\rm ({\it cf}. Figure 4)}. 
\end{prop}

\begin{figure}[ht]
  \begin{center}
    \includegraphics[height=5cm]{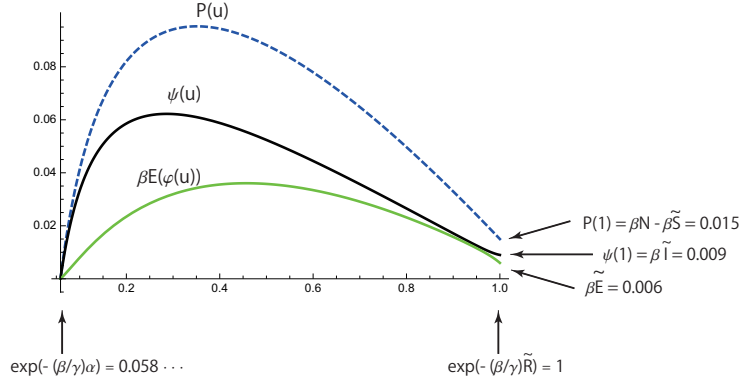}
    \caption{Variations of $P(u) = \beta N-\beta \tilde{S}
    e^{(\beta/\gamma)\tilde{R}}u+\gamma \log u$ (dashed curve), 
    $\beta E(\varphi(u))$ (green curve), and $\psi(u)$ (solid curve) 
    obtained by the numerical integration of the initial 
    value problem (\ref{ny43}), (\ref{ny44}) 
    for $N=1000, \tilde{S}=950, \tilde{E}=20, 
    \tilde{I}=30,  \tilde{R}=0, \beta=0.3/1000, \gamma=0.1$ and 
    $\delta=0.2$. In this case we get $\alpha=944.05\cdots$, 
    $e^{-(\beta/\gamma)\alpha} = 0.058\cdots$, 
    $P(1)=\beta N-\beta\tilde{S}=0.015$, 
    $\psi(1)=\beta \tilde{I}=0.009$ and 
    $\beta E(\varphi(1)) = \beta \tilde{E} = 0.006$. 
    Moreover, $\lim_{u \to e^{-(\beta/\gamma)\alpha}+0} P(u)=0$, 
    $\lim_{u \to e^{-(\beta/\gamma)\alpha}+0} \psi(u)=0$, and 
    $\lim_{u \to e^{-(\beta/\gamma)\alpha}+0} 
    \beta E(\varphi(u))=\lim_{u \to e^{-(\beta/\gamma)\alpha}+0} 
    \bigl(P(u)-\psi(u)\bigr)=0$. }
  \end{center}
\end{figure}

{\Proof}
Since 
\begin{eqnarray*}
   &   & \lim_{u \to e^{-(\beta/\gamma)\alpha}+0} 
         \bigl(\beta N-\beta\tilde{S}e^{(\beta/\gamma)\tilde{R}}u 
         +\gamma \log u \bigr)  \\
   & = & \lim_{x \to \alpha-0} 
         \beta \bigl(N - \tilde{S}e^{(\beta/\gamma)\tilde{R}}
         e^{-(\beta/\gamma)x} - x \bigr) = 0 
\end{eqnarray*}
by Lemma \ref{ny:lem6}, it follows from Proposition \ref{ny:prop1} that 
(\ref{ny59}) and (\ref{ny60}) hold by taking the limit as 
$u \to e^{-(\beta/\gamma)\alpha}+0$ in (\ref{ny56}) and (\ref{ny57}). 
\qed

\begin{lem}  \label{ny:lem10} 
Let $\psi(u)$ be the unique positive solution of the initial value 
problem {\rm (\ref{ny43}), (\ref{ny44})}. 
Then there exists the inverse function 
$\varphi^{-1}(t) \in C^{1}(0,\infty)$ of the function 
\begin{equation}
   t = \varphi(u) = \int_{u}^{e^{-(\beta/\gamma)\tilde{R}}} 
                     \frac{d\xi}{\xi\psi(\xi)} 
                     \label{ny61}
\end{equation}
for $e^{-(\beta/\gamma)\alpha} < u \leq e^{-(\beta/\gamma)\tilde{R}}$, 
such that 
$\varphi^{-1}(t)$ is decreasing on $[0,\infty)$, 
$\varphi^{-1}(0) = e^{-(\beta/\gamma)\tilde{R}}$ 
and $\lim_{t \to \infty} \varphi^{-1}(t) = e^{-(\beta/\gamma)\alpha}$. 
\end{lem}

{\Proof}
It is easy to see that $\varphi(u) \in 
C^{1}(e^{-(\beta/\gamma)\alpha},e^{-(\beta/\gamma)\tilde{R}})$, 
$\varphi(u)$ is decreasing in 
$(e^{-(\beta/\gamma)\alpha}, e^{-(\beta/\gamma)\tilde{R}}]$ and 
$\varphi\bigl(e^{-(\beta/\gamma)\tilde{R}}\bigr) = 0$. 
Dividing (\ref{ny43}) by $(\gamma+\delta)\psi(u)^{2}$ yields 
\begin{equation}
   \frac{1}{u\psi(u)} 
   = \frac{\delta}{\gamma+\delta}
   \frac{\beta N-\beta\tilde{S}e^{(\beta/\gamma)\tilde{R}}u+\gamma \log u}
        {u\psi(u)^{2}} 
   + \frac{1}{\gamma+\delta}\frac{\psi'(u)}{\psi(u)}, 
   \label{ny62}
\end{equation}
and hence
\begin{eqnarray}
   \varphi(u)
   & = & \int_{u}^{e^{-(\beta/\gamma)\tilde{R}}} \frac{d\xi}{\xi\psi(\xi)} 
         \nonumber \\
   & = & \frac{\delta}{\gamma+\delta} \int_{u}^{e^{-(\beta/\gamma)\tilde{R}}} 
         \frac{\beta N-\beta\tilde{S}e^{(\beta/\gamma)\tilde{R}}\xi 
         +\gamma \log \xi}{\xi\psi(\xi)^{2}} d\xi 
          \nonumber \\
   &   & \hspace*{30ex} + \frac{1}{\gamma+\delta} 
         \int_{u}^{e^{-(\beta/\gamma)\tilde{R}}}
         \frac{\psi'(\xi)}{\psi(\xi)}\,d\xi  
         \nonumber \\
   & = & \frac{\delta}{\gamma+\delta} \int_{u}^{e^{-(\beta/\gamma)\tilde{R}}} 
         \frac{\beta N-\beta\tilde{S}e^{(\beta/\gamma)\tilde{R}}\xi 
         +\gamma \log \xi}{\xi\psi(\xi)^{2}} d\xi 
         \nonumber \\
   &   & \hspace*{30ex}  + \frac{1}{\gamma+\delta}
           \bigl(\log\,(\beta\tilde{I}) - \log \psi(u)\bigr) 
         \nonumber \\ 
   & \geq & \frac{1}{\gamma+\delta}
           \bigl(\log\,(\beta\tilde{I}) - \log \psi(u)\bigr). 
         \label{ny63}
\end{eqnarray}
Taking account of (\ref{ny59}), we conclude that 
$\lim_{u \to e^{-(\beta/\gamma)\alpha}+0} \log \psi(u) = -\infty$, 
and hence 
$\lim_{u \to e^{-(\beta/\gamma)\alpha}+0} \varphi(u) = \infty$ 
by taking the limit as $u \to e^{-(\beta/\gamma)\alpha}+0$ 
in (\ref{ny63}). 
Therefore there exists the inverse function $\varphi^{-1}(t)$ 
which has the desired properties. 
\qed

Now we state our main theorem.

\begin{thm} \label{ny:thm4}
The function $(S(t),E(t),I(t),R(t))$ defined by 
\begin{eqnarray}
   S(t) & = & \tilde{S}e^{(\beta/\gamma)\tilde{R}}\varphi^{-1}(t), 
              \label{ny64} \\
   E(t) & = & \tilde{E}e^{-\delta t} 
              + \tilde{S}e^{(\beta/\gamma)\tilde{R}}e^{-\delta t}
              \int_{\varphi^{-1}(t)}^{e^{-(\beta/\gamma)\tilde{R}}}
              e^{\delta\varphi(v)}dv, 
              \label{ny65} \\
   I(t) & = & N - \tilde{S}e^{(\beta/\gamma)\tilde{R}}\varphi^{-1}(t) 
              + \frac{\gamma}{\beta}\log \varphi^{-1}(t) 
              - \tilde{E}e^{-\delta t}  \nonumber\\
        &   &  \qquad - \tilde{S}e^{(\beta/\gamma)\tilde{R}}e^{-\delta t}
              \int_{\varphi^{-1}(t)}^{e^{-(\beta/\gamma)\tilde{R}}}
              e^{\delta\varphi(v)}dv,   
              \label{ny66} \\
   R(t) & = & - \frac{\gamma}{\beta}\,\log \varphi^{-1}(t) 
              \label{ny67}
\end{eqnarray}
is a solution of the initial value problem {\rm (\ref{ny1})--(\ref{ny5})}, 
where $\varphi(u)$ and $\varphi^{-1}(t)$ are given in 
Lemma {\rm \ref{ny:lem10}}. 
\end{thm}

{\Proof} 
First we note that 
\begin{eqnarray}
   \bigl(\varphi^{-1}(t)\bigr)'
   & = & \frac{1}{\varphi'(u)}\Big\vert_{u=\varphi^{-1}(t)} 
         = - u\psi(u)\Big\vert_{u=\varphi^{-1}(t)} 
         \nonumber \\
   & = & - \varphi^{-1}(t)\psi\bigl(\varphi^{-1}(t)\bigr) 
         = - \beta \varphi^{-1}(t) I(t) 
         \label{ny68}
\end{eqnarray}
by means of (\ref{ny46}) and (\ref{ny66}). 
It follows from (\ref{ny64}) and (\ref{ny68}) that 
\begin{eqnarray}
   S'(t) 
   & = & \tilde{S}e^{(\beta/\gamma)\tilde{R}}\bigl(\varphi^{-1}(t)\bigr)' 
         \nonumber \\
   & = & - \beta \tilde{S}e^{(\beta/\gamma)\tilde{R}}\varphi^{-1}(t)I(t) 
         \nonumber \\
   & = & - \beta S(t)I(t)
         \label{ny69}
\end{eqnarray}
and therefore (\ref{ny1}) holds. 
A direct calculation gives 
\begin{eqnarray}
    E'(t) 
   & = & - \delta\tilde{E}e^{-\delta t}  \nonumber \\
   &   & + \tilde{S}e^{(\beta/\gamma)\tilde{R}}
         \left(-\delta e^{-\delta t} 
         \int_{\varphi^{-1}(t)}^{e^{-(\beta/\gamma)\tilde{R}}}
          e^{\delta\varphi(v)}dv 
         + e^{-\delta t}\bigl(-e^{\delta t}\bigl(\varphi^{-1}(t)\bigr)'\bigr)
         \right)  \nonumber \\
   & = & - \delta\tilde{E}e^{-\delta t} 
         - \delta \tilde{S}e^{(\beta/\gamma)\tilde{R}}
         e^{-\delta t}\int_{\varphi^{-1}(t)}^{e^{-(\beta/\gamma)\tilde{R}}}
          e^{\delta\varphi(v)}dv  
         - \tilde{S}e^{(\beta/\gamma)\tilde{R}}\bigl(\varphi^{-1}(t)\bigr)' 
         \nonumber \\
   & = & - \delta E(t) + \beta S(t)I(t) 
         \label{ny70} 
\end{eqnarray}
in view of (\ref{ny65}) and (\ref{ny69}), and hence (\ref{ny2}) holds. 
A simple computation shows that 
\begin{eqnarray}
   I'(t) 
   & = & -\tilde{S}e^{(\beta/\gamma)\tilde{R}}\bigl(\varphi^{-1}(t)\bigr)' 
         + \frac{\gamma}{\beta}\,\frac{\bigl(\varphi^{-1}(t)\bigr)'}
           {\varphi^{-1}(t)}   \nonumber \\
   &   & \quad - \left(\tilde{E}e^{-\delta t} 
         + \tilde{S}e^{(\beta/\gamma)\tilde{R}}e^{-\delta t}
         \int_{\varphi^{-1}(t)}^{e^{-(\beta/\gamma)\tilde{R}}}
          e^{\delta\varphi(v)}dv \right)'   
         \label{ny71} \\
   & = & \beta S(t)I(t) + \frac{\gamma}{\beta}\bigl(- \beta I(t)\bigr) 
         - E'(t)   \nonumber \\
   & = & \beta S(t)I(t) - \gamma I(t) 
         - \bigl( -\delta E(t) + \beta S(t)I(t)\bigr)   \nonumber \\
   & = & \delta E(t) - \gamma I(t)  \nonumber
\end{eqnarray}
by taking account of (\ref{ny68})--(\ref{ny70}). 
Thus we find that (\ref{ny3}) holds. 
It is easy to verify from (\ref{ny68}) that 
$$
   R'(t) 
    =  - \frac{\gamma}{\beta} 
         \frac{\left(\varphi^{-1}(t)\right)'}{\varphi^{-1}(t)} 
         =  - \frac{\gamma}{\beta}(- \beta I(t)) 
         =  \gamma I(t) 
$$
which is the desired equation (\ref{ny4}). 
We easily see that 
\begin{eqnarray*}
   S(0) 
   & = & \tilde{S}e^{(\beta/\gamma)\tilde{R}}\varphi^{-1}(0) 
         = \tilde{S}e^{(\beta/\gamma)\tilde{R}}e^{-(\beta/\gamma)\tilde{R}} 
         = \tilde{S}, \\
   E(0) 
   & = & \tilde{E} + \tilde{S}e^{(\beta/\gamma)\tilde{R}}
              \int_{\varphi^{-1}(0)}^{e^{-(\beta/\gamma)\tilde{R}}}
              e^{\delta\varphi(v)}dv 
         = \tilde{E}, \\
   I(0) 
   & = & N - \tilde{S} 
         + \frac{\gamma}{\beta}\left(- \frac{\beta}{\gamma}\tilde{R}\right) 
         - \tilde{E} 
         = \tilde{I}, \\
   R(0) 
   & = & - \frac{\gamma}{\beta}\left(- \frac{\beta}{\gamma}\tilde{R}\right) 
         = \tilde{R} 
\end{eqnarray*}
in light of $\varphi^{-1}(0) = e^{-(\beta/\gamma)\tilde{R}}$. 
\qed

\begin{thm} \label{ny:thm5}
Let $(S(t), E(t), I(t), R(t))$ be the exact solution 
{\rm (\ref{ny64})--(\ref{ny67})} of the initial value problem 
{\rm (\ref{ny1})--(\ref{ny5})}, and let 
$$
   \bigl(\hat{S}(u),\hat{E}(u),\hat{I}(u),\hat{R}(u)\bigr) 
   := \bigl(S(\varphi(u)),E(\varphi(u)),I(\varphi(u)),R(\varphi(u))\bigr). 
$$
Then, 
$\bigl(\hat{S}(u),\hat{E}(u),\hat{I}(u),\hat{R}(u)\bigr)$ satisfies 
the linear differential system 
\begin{eqnarray}
   & & \frac{d\hat{S}(u)}{du}  =  \frac{\hat{S}(u)}{u}, 
       \label{ny72} \\
   & & \frac{d\hat{E}(u)}{du} - \frac{\delta}{u\psi(u)}\hat{E}(u) 
       = - \frac{\hat{S}(u)}{u}, 
       \label{ny73} \\ 
   & & \frac{d\hat{I}(u)}{du} 
       - \frac{\gamma}{\beta}\frac{1}{u} 
       = - \frac{\delta}{u\psi(u)} \hat{E}(u), 
       \label{ny74} \\ 
   & & \frac{d\hat{R}(u)}{du}  =  - \frac{\gamma}{\beta}\,\frac{1}{u}
       \label{ny75} 
\end{eqnarray}
for $u \in (e^{-(\beta/\gamma)\alpha}, e^{-(\beta/\gamma)\tilde{R}})$, 
and the initial condition
\begin{eqnarray}
   \hat{S}\left(e^{-(\beta/\gamma)\tilde{R}}\right) & = & \tilde{S}, 
                                \label{ny76}  \\
   \hat{E}\left(e^{-(\beta/\gamma)\tilde{R}}\right) & = & \tilde{E}, 
                                \label{ny77}  \\
   \hat{I}\left(e^{-(\beta/\gamma)\tilde{R}}\right) & = & \tilde{I}, 
                                \label{ny78}  \\
   \hat{R}\left(e^{-(\beta/\gamma)\tilde{R}}\right) & = & \tilde{R}. 
                                \label{ny79} 
\end{eqnarray}
\end{thm}

{\Proof}
We see from (\ref{ny68}) that 
\begin{equation}
   \hat{I}(u) = I(\varphi(u)) = \frac{1}{\beta}\psi(u). 
   \label{ny80}
\end{equation}
Since $S(t)$ satisfies (\ref{ny1}), we find that
$$
   S'(\varphi(u)) = - \beta S(\varphi(u))I(\varphi(u)) 
                  = - \beta \hat{S}(u)\hat{I}(u). 
$$
We obtain 
\begin{eqnarray*}
   \frac{d\hat{S}(u)}{du}
   & = & \frac{dS(t)}{dt}\Big\vert_{t=\varphi(u)}\varphi'(u)
         =  S'(\varphi(u))\left(- \frac{1}{u\psi(u)}\right) \\
   & = & \left(- \beta \hat{S}(u)\hat{I}(u)\right)
         \left(- \frac{1}{u\psi(u)}\right) \\
   & = & \frac{\hat{S}(u)}{u}
\end{eqnarray*}
by taking account of (\ref{ny80}), and therefore (\ref{ny72}) follows. 
We observe, using (\ref{ny2}) and (\ref{ny80}), that 
\begin{eqnarray*}
   \frac{d\hat{E}(u)}{du}
   & = & \frac{dE(t)}{dt}\Big\vert_{t=\varphi(u)}\varphi'(u)
         =  E'(\varphi(u))\left(- \frac{1}{u\psi(u)}\right) \\
   & = & \left(\beta \hat{S}(u)\hat{I}(u) - \delta \hat{E}(u) \right)
         \left(- \frac{1}{u\psi(u)}\right) \\
   & = & - \frac{\hat{S}(u)}{u} + \frac{\delta}{u\psi(u)}\hat{E}(u), 
\end{eqnarray*}
which is the desired equation (\ref{ny73}). 
Using (\ref{ny3}), we obtain 
\begin{eqnarray*}
   \frac{d\hat{I}(u)}{du} 
   & = & \frac{dI(t)}{dt}\Big\vert_{t=\varphi(u)}\varphi'(u) 
         \\
   & = & \bigl(\delta \hat{E}(u) - \gamma \hat{I}(u)\bigr) 
         \left(- \frac{1}{u\psi(u)} \right) 
         \\
   & = & - \delta \frac{\hat{E}(u)}{u\psi(u)} 
         + \gamma \frac{\hat{I}(u)}{u\psi(u)} 
         \\
   & = & -  \frac{\delta}{u\psi(u)} \hat{E}(u)
         + \frac{\gamma}{\beta} \frac{1}{u}, 
\end{eqnarray*}
and hence (\ref{ny74}) holds. 
Employing (\ref{ny4}) and (\ref{ny80}) yields 
\begin{eqnarray*}
   \frac{d\hat{R}(u)}{du} & = & 
         \frac{dR(t)}{dt}\Big\vert_{t=\varphi(u)}\varphi'(u)
          =  R'(\varphi(u))\left(- \frac{1}{u\psi(u)}\right) \\
   & = & \gamma \hat{I}(u) \left(- \frac{1}{u\psi(u)}\right) \\
   & = & - \frac{\gamma}{\beta} \frac{1}{u}, 
\end{eqnarray*}
which is equal to (\ref{ny75}). 
It is obvious that
$$
   \hat{S}\left(e^{-(\beta/\gamma)\tilde{R}}\right) 
   =  S\left(\varphi\left(e^{-(\beta/\gamma)\tilde{R}}\right)\right) \\  
   =  S(0) = \tilde{S}, 
$$
$$
   \hat{E}\left(e^{-(\beta/\gamma)\tilde{R}}\right) 
   =  E\left(\varphi\left(e^{-(\beta/\gamma)\tilde{R}}\right)\right) \\  
   =  E(0) = \tilde{E}, 
$$
$$
   \hat{I}\left(e^{-(\beta/\gamma)\tilde{R}}\right) 
   =  I\left(\varphi\left(e^{-(\beta/\gamma)\tilde{R}}\right)\right) \\  
   =  I(0) = \tilde{I}, 
$$
$$
   \hat{R}\left(e^{-(\beta/\gamma)\tilde{R}}\right) 
   =  R\left(\varphi\left(e^{-(\beta/\gamma)\tilde{R}}\right)\right) \\  
   =  R(0) = \tilde{R}, 
$$
and therefore (\ref{ny76})--(\ref{ny79}) are satisfied. 
\qed

\begin{thm} \label{ny:thm6}
Solving the initial value problem {\rm (\ref{ny72})--(\ref{ny79})}, 
we obtain the parametric solution {\rm (\ref{ny9})--(\ref{ny12})} 
for $e^{-(\beta/\gamma)\alpha} < u \leq e^{-(\beta/\gamma)\tilde{R}}$. 
\end{thm}

{\Proof}
Since (\ref{ny72}) is equivalent to 
$$
   \frac{d}{du}\left(\frac{1}{u}\hat{S}(u)\right) = 0, 
$$
we obtain 
$$
   \hat{S}(u) = ku 
$$
for some constant $k$.  It follows from (\ref{ny76}) that
$$
   \hat{S}\left(e^{-(\beta/\gamma)\tilde{R}}\right) 
   = ke^{-(\beta/\gamma)\tilde{R}} = \tilde{S} 
$$
which implies 
$$
   k = \tilde{S}e^{(\beta/\gamma)\tilde{R}}. 
$$
Hence we get 
\begin{equation}
   \hat{S}(u) = \tilde{S}e^{(\beta/\gamma)\tilde{R}}u. 
   \label{ny81}
\end{equation}
From (\ref{ny81}) we see that 
$$
   - \frac{\hat{S}(u)}{u} = - \tilde{S}e^{(\beta/\gamma)\tilde{R}} 
$$
and therefore (\ref{ny73}) reduces to 
\begin{equation}
   \frac{d\hat{E}(u)}{du} - \frac{\delta}{u\psi(u)}\hat{E}(u) 
       = - \tilde{S}e^{(\beta/\gamma)\tilde{R}} 
         \label{ny82}
\end{equation}
which is written in the form 
\begin{equation}
   \frac{d\hat{E}(u)}{du} + \delta \varphi'(u)\hat{E}(u) 
   = - \tilde{S}e^{(\beta/\gamma)\tilde{R}}. 
      \label{ny83}
\end{equation}
Multiplying (\ref{ny83}) by $e^{\delta \varphi(u)}$ yields
$$
   \frac{d}{du}\left(e^{\delta\varphi(u)}\hat{E}(u)\right) 
   = - \tilde{S}e^{(\beta/\gamma)\tilde{R}}e^{\delta\varphi(u)}, 
$$
and integrating the above on $[u, e^{-(\beta/\gamma)\tilde{R}}]$ gives 
$$
   \hat{E}(u) = e^{-\delta\varphi(u)}
   \left(\tilde{E} + \tilde{S}e^{(\beta/\gamma)\tilde{R}}
   \int_{u}^{e^{-(\beta/\gamma)\tilde{R}}} e^{\delta\varphi(v)}dv\right) 
$$
which is equal to (\ref{ny10}). 
Multiplying (\ref{ny74}) by $\beta$, we obtain 
\begin{equation}
   \frac{d(\beta\hat{I}(u))}{du} 
   - \frac{\gamma}{u}  
   = - \frac{\beta\delta}{u\psi(u)}\hat{E}(u). 
   \label{ny84}
\end{equation}
We define $z(u)$ by 
$$
   z(u) := \beta\hat{I}(u) 
         - \bigl(\beta N - \beta\tilde{S}e^{(\beta/\gamma)\tilde{R}}u 
         + \gamma \log u \bigr) 
$$
to get 
\begin{equation}
      \frac{dz(u)}{du}  = \frac{d(\beta\hat{I}(u))}{du} 
      + \beta\tilde{S}e^{(\beta/\gamma)\tilde{R}} - \frac{\gamma}{u}. 
      \label{ny85}
\end{equation}
Combining (\ref{ny84}) with (\ref{ny85}) yields 
\begin{eqnarray}
   \frac{dz(u)}{du} 
   & = & \beta\tilde{S}e^{(\beta/\gamma)\tilde{R}} 
         - \frac{\beta\delta}{u\psi(u)} \hat{E}(u) 
         \nonumber \\
   & = & -\beta\left( - \tilde{S}e^{(\beta/\gamma)\tilde{R}} 
         + \frac{\delta}{u\psi(u)} \hat{E}(u) \right). 
         \label{ny86}
\end{eqnarray}
It follows from (\ref{ny82}) and (\ref{ny86}) that 
$$
   \frac{dz(u)}{du}  = - \beta \frac{d\hat{E}(u)}{du}, 
$$
and hence 
$$
   z(u) = - \beta \hat{E}(u) + k 
$$
for some constant $k$. Since 
\begin{eqnarray*}
   z\bigl(e^{-(\beta/\gamma)\tilde{R}}\bigr) 
   & = & \beta \hat{I}\bigl(e^{-(\beta/\gamma)\tilde{R}}\bigr) 
         - \bigl(\beta N - \beta\tilde{S} - \beta\tilde{R}\bigr) \\
   & = & \beta\tilde{I} 
         - \bigl(\beta N - \beta\tilde{S} - \beta\tilde{R}\bigr) 
         = - \beta\tilde{E} 
\end{eqnarray*}
and 
$- \beta\hat{E}\bigl(e^{-(\beta/\gamma)\tilde{R}}\bigr) = - \beta\tilde{E}$, 
we find that $k = 0$, and hence $z(u) = - \beta\hat{E}(u)$, i.e., 
$$
   \beta\hat{I}(u) = 
         \bigl(\beta N - \beta\tilde{S}e^{(\beta/\gamma)\tilde{R}}u 
         + \gamma \log u \bigr) - \beta\hat{E}(u), 
$$
which is equivalent to (\ref{ny11}). 
Solving (\ref{ny75}) yields 
$$
   \hat{R}(u) = - \frac{\gamma}{\beta}\log u + k
$$
for some constant $k$. The initial condition (\ref{ny79}) implies 
$$
   \hat{R}\left(e^{-(\beta/\gamma)\tilde{R}}\right) 
   = - \frac{\gamma}{\beta}\log e^{-(\beta/\gamma)\tilde{R}} + k 
   = \tilde{R} + k = \tilde{R} 
$$
and hence $k = 0$. Consequently we have
$$
   \hat{R}(u) = - \frac{\gamma}{\beta}\log u.
$$
\qed

\begin{rem} \label{ny:rem4} \rm 
Let $I(t)$ be given by (\ref{ny66}). Then $I(t)$ can be written in the 
simple form 
$$
   I(t) = \frac{1}{\beta}\psi\bigl(\varphi^{-1}(t)\bigr) 
$$
in view of (\ref{ny46}) and (\ref{ny66}). 
\end{rem}


\section{Various properties of solution}

In this section we obtain various properties of solution by 
investigating the exact solution of the initial value 
problem (\ref{ny1})--(\ref{ny5}). 

\begin{thm} \label{ny:thm7}
Let $R(t)$ be given by {\rm (\ref{ny67})}. Then 
we observe that $R(\infty) = \alpha$, 
\begin{equation}
   R(\infty) = N - \tilde{S}e^{(\beta/\gamma)\tilde{R}}
       e^{-(\beta/\gamma)R(\infty)},  \label{ny87}
\end{equation}
and that $R(t)$ is an increasing function on $[0,\infty)$ such that 
$$
   \tilde{R} \leq R(t) < \alpha = R(\infty). 
$$
\end{thm}

{\Proof}
It can be shown that 
\begin{eqnarray*}
   R(\infty) = \lim_{t \to \infty} R(t) 
   & = & \lim_{t \to \infty} 
         - \frac{\gamma}{\beta}\log\,\varphi^{-1}(t) \\
   & = & \lim_{u \to e^{-(\beta/\gamma)\alpha}+0} 
         - \frac{\gamma}{\beta}\log\,u  \\
   & = & \alpha.
\end{eqnarray*}
Since $\alpha = R(\infty)$, the identity (\ref{ny87}) 
follows from the definition of $\alpha$ (see Lemma \ref{ny:lem6}). 
Taking account of 
$e^{-(\beta/\gamma)\alpha} < \varphi^{-1}(t) \leq e^{-(\beta/\gamma)\tilde{R}}$, 
we find that 
$$
   - \frac{\gamma}{\beta} \log\,e^{-(\beta/\gamma)\tilde{R}} 
   \leq R(t) < - \frac{\gamma}{\beta} \log\,e^{-(\beta/\gamma)\alpha} 
$$
or
$$
   \tilde{R} \leq R(t) < \alpha = R(\infty). 
$$
It is readily seen that $R(t)$ is increasing on $[0,\infty)$ 
in view of the fact that 
$\varphi^{-1}(t)$ is decreasing on $[0,\infty)$. 
\qed

\begin{thm} \label{ny:thm8}
Let $S(t)$ be given by {\rm (\ref{ny64})}. Then we see that
\begin{equation}
   S(\infty) = \tilde{S}e^{(\beta/\gamma)\tilde{R}}
       e^{-(\beta/\gamma)R(\infty)},  \label{ny88}
\end{equation}
and that $S(t)$ is a deceasing function on $[0,\infty)$ such that 
$$
       \tilde{S} \geq S(t) > \tilde{S}e^{(\beta/\gamma)\tilde{R}}
       e^{-(\beta/\gamma)\alpha} 
       = S(\infty).
$$
\end{thm}

{\Proof}
The identity (\ref{ny88}) follows from 
\begin{eqnarray*}
   S(\infty) = \lim_{t \to \infty} S(t) 
   & = & \lim_{t \to \infty} 
         \tilde{S}e^{(\beta/\gamma)\tilde{R}}\varphi^{-1}(t) \\
   & = & \lim_{u \to e^{-(\beta/\gamma)\alpha}+0} 
         \tilde{S}e^{(\beta/\gamma)\tilde{R}}u \\
   & = & \tilde{S}e^{(\beta/\gamma)\tilde{R}}
       e^{-(\beta/\gamma)\alpha} \\
   & = & \tilde{S}e^{(\beta/\gamma)\tilde{R}}
       e^{-(\beta/\gamma)R(\infty)}. 
\end{eqnarray*}
Since 
$e^{-(\beta/\gamma)\alpha} < \varphi^{-1}(t) \leq e^{-(\beta/\gamma)\tilde{R}}$, we get 
$$
   \tilde{S}e^{(\beta/\gamma)\tilde{R}}
       e^{-(\beta/\gamma)\alpha} 
   < \tilde{S}e^{(\beta/\gamma)\tilde{R}}\varphi^{-1}(t) 
   \leq \tilde{S}e^{(\beta/\gamma)\tilde{R}}
       e^{-(\beta/\gamma)\tilde{R}}. 
$$
Hence we obtain 
$$
   \tilde{S}e^{(\beta/\gamma)\tilde{R}}
       e^{-(\beta/\gamma)\alpha}
   < S(t) \leq \tilde{S}. 
$$
Since $\varphi^{-1}(t)$ is decreasing on $[0,\infty)$, we conclude that 
$S(t)$ is also decreasing on $[0,\infty)$. 
\qed

\begin{thm} \label{ny:thm9}
Let $E(t)$ be given by {\rm (\ref{ny65})}. Then we find that 
\begin{eqnarray*}
   & & E(\infty) = 0, \\
   & & E(t) > 0 \quad \mbox{on}\ [0,\infty), 
\end{eqnarray*}
and $E(t)$ has the maximum $\max_{t \geq 0} E(t)$ at some 
$t = T_{1} \in \{T;\, E'(T)=0\}$, where 
\begin{eqnarray*}
   E'(T) 
    & = & \left(\frac{\delta}{\beta} 
         + \tilde{S}e^{(\beta/\gamma)\tilde{R}}\varphi^{-1}(T)\right)
         \psi\bigl(\varphi^{-1}(T)\bigr)  \\
    &   & - \delta \left( N - \tilde{S}e^{(\beta/\gamma)\tilde{R}}
         \varphi^{-1}(T) 
              + \frac{\gamma}{\beta}\log \varphi^{-1}(T)\right). 
\end{eqnarray*}
\end{thm}

{\Proof}
It can be shown that
\begin{eqnarray*}
   E(\infty)
   & = & \lim_{t \to \infty} E(t) \\
   & = & \lim_{u \to e^{-(\beta/\gamma)\alpha}+0} 
         \left( \tilde{E}e^{-\delta\varphi(u)} 
           + \tilde{S}e^{(\beta/\gamma)\tilde{R}}e^{-\delta\varphi(u)}
           \int_{u}^{e^{-(\beta/\gamma)\tilde{R}}}
           e^{\delta\varphi(v)}dv \right) \\
   & = & 0
\end{eqnarray*}
by means of (\ref{ny60}) in Proposition \ref{ny:prop2}. 
Since $e^{-(\beta/\gamma)\alpha} < \varphi^{-1}(t) \leq 
e^{-(\beta/\gamma)\tilde{R}}$\ $(t \geq 0)$ and 
 $\hat{E}(u) > 0$ 
for $e^{-(\beta/\gamma)\alpha} < u \leq e^{-(\beta/\gamma)\tilde{R}}$ 
(cf. (\ref{ny58})), we easily check that  
$E(t) = \hat{E}\bigl(\varphi^{-1}(t)\bigr) > 0$ on $[0,\infty)$. 
The hypothesis (A$_{3}$) implies that the right differential derivative 
$E_{+}'(0)$ is positive because 
$$
   E_{+}'(0) = \lim_{t \to +0} E'(t) 
   = \lim_{t \to +0} \bigl(\beta S(t)I(t) - \delta E(t)\bigr) 
   = \beta \tilde{S}\tilde{I} - \delta \tilde{E} > 0. 
$$
Since the definition of $E_{+}'(0)$ means 
$$
   0 < E_{+}'(0) = \lim_{t \to +0} \frac{E(t) - E(0)}{t} 
   = \lim_{t \to +0} \frac{E(t) - \tilde{E}}{t}, 
$$
we find that 
for $\varepsilon = (1/2) E_{+}'(0) > 0$ there exists a number 
$\delta_{\varepsilon} > 0$ 
such that 
$$
   \left\vert \frac{E(t) - \tilde{E}}{t} - E_{+}'(0) \right\vert 
   < \frac{1}{2} E_{+}'(0) 
$$
holds for $0 < t < \delta_{\varepsilon}$, and therefore 
$$
   \frac{1}{2} E_{+}'(0) < \frac{E(t) - \tilde{E}}{t} 
$$
or 
$$
   E(t) > \tilde{E} + \frac{1}{2} E_{+}'(0) t > \tilde{E}
$$
holds for $0 < t < \delta_{\varepsilon}$. 
Since $E(\infty) = 0$, there exists a number $\tilde{T}$ such that 
$E(\tilde{T}) = \tilde{E}$ and $E(t) \leq \tilde{E}$ for $t \geq \tilde{T}$. 
Therefore there exists $\max_{0 \leq t \leq \tilde{T}} E(t) 
= E(T_{1})\ ( > \tilde{E})$ at some $t = T_{1}\ (< \tilde{T})$. 
Since $E(t) \leq \tilde{E}$ for $t \geq \tilde{T}$, 
we find that $\max_{t \geq 0} E(t) = \max_{0 \leq t \leq \tilde{T}} E(t) 
= E(T_{1})$. 
It is obvious that $E'(T_{1}) = 0$. 
It is easy to check from (\ref{ny64})--(\ref{ny66}) and (\ref{ny80}) that 
\begin{eqnarray}
   E'(t) 
   & = & - \delta E(t) + \beta S(t)I(t)  \nonumber\\
   & = & - \delta E(t) 
         + \tilde{S}e^{(\beta/\gamma)\tilde{R}}\varphi^{-1}(t)
         \psi\bigl(\varphi^{-1}(t)\bigr)  \nonumber \\
   & = & - \delta \left( N - \tilde{S}e^{(\beta/\gamma)\tilde{R}}
         \varphi^{-1}(t) 
              + \frac{\gamma}{\beta}\log \varphi^{-1}(t) 
         - \frac{1}{\beta}\psi\bigl(\varphi^{-1}(t)\bigr)\right) 
         \nonumber \\
   &   & \quad + \tilde{S}e^{(\beta/\gamma)\tilde{R}}\varphi^{-1}(t)
         \psi\bigl(\varphi^{-1}(t)\bigr)  \nonumber \\
   & = & \left(\frac{\delta}{\beta} 
         + \tilde{S}e^{(\beta/\gamma)\tilde{R}}\varphi^{-1}(t)\right)
         \psi\bigl(\varphi^{-1}(t)\bigr)  \nonumber \\
   &   & \quad - \delta \left( N - \tilde{S}e^{(\beta/\gamma)\tilde{R}}
         \varphi^{-1}(t) 
              + \frac{\gamma}{\beta}\log \varphi^{-1}(t)\right). 
         \label{ny89}
\end{eqnarray}
\qed

\begin{rem} \label{ny:rem5} \rm 
If $u_{1}$ is a unique solution of the equation 
$$
   \left(\frac{\delta}{\beta} 
         + \tilde{S}e^{(\beta/\gamma)\tilde{R}}u \right)
         \psi(u)  
    =  \delta \left( N - \tilde{S}e^{(\beta/\gamma)\tilde{R}}u 
              + \frac{\gamma}{\beta}\log u \right), 
$$
then we obtain
$$
   T_{1} = \varphi(u_{1}) 
         = \int_{u_{1}}^{e^{-(\beta/\gamma)\tilde{R}}} 
           \frac{d\xi}{\xi\psi(\xi)}
$$
by means of (\ref{ny61}) {\rm ({\it cf}. Figure 5)}.

\begin{figure}[ht]
  \begin{center}
    \includegraphics[height=5cm]{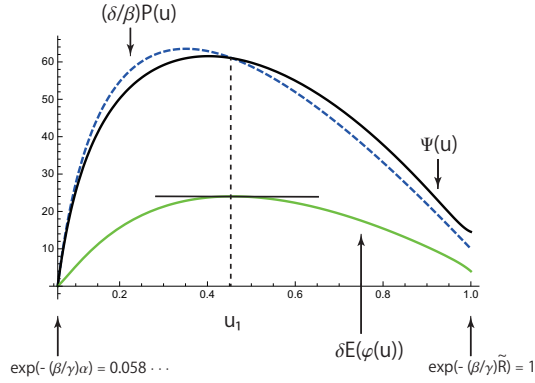}
    \caption{Variations of $(\delta/\beta)P(u) = \delta(N-\tilde{S}
    e^{(\beta/\gamma)\tilde{R}}u+(\gamma/\beta) \log u)$ (dashed curve), 
    $\delta E(\varphi(u))$ (green curve) and $\Psi(u)$ (solid curve) 
    obtained by the numerical integration of the initial value problem 
    (\ref{ny90}), (\ref{ny91}) 
    for $N=1000, \tilde{S}=950, \tilde{E}=20, 
    \tilde{I}=30,  \tilde{R}=0, \beta=0.3/1000, \gamma=0.1$ and 
    $\delta=0.2$. In this case we find that there exists a unique $u_{1}$ 
    such that $(\delta/\beta)P(u_{1}) 
    = \Psi(u_{1})$, and that $T_{1}$ is calculated by }
\end{center}
$$
   T_{1} = \varphi(u_{1}) = \int_{u_{1}}^{1} 
           \frac{d\xi}{\xi\psi(\xi)}, 
$$
where $\psi(u)$ is a unique positive solution of the initial value problem 
\begin{eqnarray*}
   &   & \psi'\psi-\frac{0.3}{u}\psi=-0.2\,\frac{0.3-0.285 u+0.1 \log u}{u} 
         \ (0.058\cdots < u < 1), \\
   &   & \psi(1) = 0.009. 
\end{eqnarray*}
\end{figure}

In case $E'(T_{1}) = 0$, we get $\beta S(T_{1})I(T_{1}) = \delta E(T_{1})$ 
by (\ref{ny2}), and hence $E(T_{1}) = (\beta/\delta)S(T_{1})I(T_{1})$. 
Therefore, in Theorem \ref{ny:thm9} we find that 
$$
   \max_{t \geq 0} E(t) = E(T_{1}) = \frac{\beta}{\delta}S(T_{1})I(T_{1}). 
$$
Letting 
$$
   \Psi(u) := 
   \left(\frac{\delta}{\beta} + \tilde{S}e^{(\beta/\gamma)\tilde{R}}u \right)
   \psi(u), 
$$
we see that $\Psi(u)$ is a solution of the initial value problem for the 
Abel differential equation
\begin{eqnarray}
   &   & \hspace*{-5ex} 
         \Psi'(u)\Psi(u) - \frac{\tilde{S}e^{(\beta/\gamma)\tilde{R}}}
         {(\delta/\beta)+ \tilde{S}e^{(\beta/\gamma)\tilde{R}}u} \Psi(u)^{2} 
          - \frac{\gamma+\delta}{u}\left(\frac{\delta}{\beta} 
         + \tilde{S}e^{(\beta/\gamma)\tilde{R}}u \right)\Psi(u) 
         \nonumber \\
   & = & - \delta\left(\frac{\delta}{\beta} 
         + \tilde{S}e^{(\beta/\gamma)\tilde{R}}u \right)^{2} 
         \frac{\beta N-\beta \tilde{S}e^{(\beta/\gamma)\tilde{R}}u 
         + \gamma \log u}{u} 
         \label{ny90}
\end{eqnarray}
for $e^{-(\beta/\gamma)\alpha} < u < e^{-(\beta/\gamma)\tilde{R}}$, 
with the initial condition 
\begin{equation}
   \Psi\bigl(e^{-(\beta/\gamma)\tilde{R}}\bigr) 
   = \beta\left(\frac{\delta}{\beta} + \tilde{S}\right)\tilde{I}. 
   \label{ny91}
\end{equation}
\end{rem}

\begin{thm} \label{ny:thm10}
Let $I(t)$ be given by {\rm (\ref{ny66})}. Then we conclude that
\begin{eqnarray*}
   & & I(\infty) = 0, \\
   & & I(t) > 0 \quad \mbox{on}\ [0,\infty), 
\end{eqnarray*}
and $I(t)$ has the maximum $\max_{t \geq 0} I(t)$ at some 
$t = T_{2} \in \{ T;\ I'(T)=0\}$, where 
$$
   I'(T) = - \frac{\gamma + \delta}{\beta} \psi\bigl(\varphi^{-1}(T)\bigr) 
           + \delta \left( N - \tilde{S}e^{(\beta/\gamma)\tilde{R}}
            \varphi^{-1}(T) 
            + \frac{\gamma}{\beta}\log \varphi^{-1}(T)\right). 
$$
\end{thm}

{\Proof}
Taking account of  (\ref{ny59}) and (\ref{ny80}) yields 
\begin{eqnarray*}
   I(\infty)
   & = & \lim_{t \to \infty} I(t) 
         = \lim_{t \to \infty} \frac{1}{\beta}\psi\bigl(\varphi^{-1}(t)\bigr)\\
   & = & \lim_{u \to e^{-(\beta/\gamma)\alpha}+0} 
         \frac{1}{\beta}\psi(u) \\
   & = & 0. 
\end{eqnarray*}
Since $e^{-(\beta/\gamma)\alpha} < \varphi^{-1}(t) \leq 
e^{-(\beta/\gamma)\tilde{R}}$\ $(t \geq 0)$ and 
$\psi(u) > 0$ for 
$e^{-(\beta/\gamma)\alpha} < u \leq e^{-(\beta/\gamma)\tilde{R}}$, 
we observe that 
$I(t) = (1/\beta)\psi\bigl(\varphi^{-1}(t)\bigr) > 0$ on $[0,\infty)$. 
The hypothesis (A$_{2}$) means that the right differential derivative 
$I_{+}'(0)$ is positive because 
$$
   I_{+}'(0) = \lim_{t \to +0} I'(t) 
   = \lim_{t \to +0} \bigl(\delta E(t) - \gamma I(t)\bigr) 
   = \delta \tilde{E} - \gamma \tilde{I} > 0, 
$$
and hence there exists a number $\delta_{1} > 0$ such that 
$I(t) > \tilde{I}$ in $(0,\delta_{1})$ as in the proof 
of Theorem \ref{ny:thm9}. 
Since $I(\infty) = 0$, we can use the same arguments as in the proof 
of Theorem \ref{ny:thm9} to conclude that there exists the maximum 
$\max_{t \geq 0} I(t) = I(T_{2})$ for some $T_{2}$. 
Then $I'(T_{2}) = 0$, and the following holds:
\begin{eqnarray*}
   I'(t)
   & = & \tilde{S}e^{(\beta/\gamma)\tilde{R}}\varphi^{-1}(t)
         \psi\bigl(\varphi^{-1}(t)\bigr) 
         - \frac{\gamma}{\beta}\psi\bigl(\varphi^{-1}(t)\bigr) 
         - E'(t) \\
   & = & \tilde{S}e^{(\beta/\gamma)\tilde{R}}\varphi^{-1}(t)
         \psi\bigl(\varphi^{-1}(t)\bigr) 
         - \frac{\gamma}{\beta}\psi\bigl(\varphi^{-1}(t)\bigr) \\
   &   & \quad - \left[ 
         \left(\frac{\delta}{\beta} 
         + \tilde{S}e^{(\beta/\gamma)\tilde{R}}\varphi^{-1}(t)\right)
         \psi\bigl(\varphi^{-1}(t)\bigr) \right. \\
   &   & \qquad \quad 
         \left. - \delta \left( N - \tilde{S}e^{(\beta/\gamma)\tilde{R}}
         \varphi^{-1}(t) 
         + \frac{\gamma}{\beta}\log \varphi^{-1}(t)\right)\right] \\
   & = & - \frac{\gamma + \delta}{\beta} \psi\bigl(\varphi^{-1}(t)\bigr) 
          + \delta \left( N - \tilde{S}e^{(\beta/\gamma)\tilde{R}}
            \varphi^{-1}(t) 
            + \frac{\gamma}{\beta}\log \varphi^{-1}(t)\right) 
\end{eqnarray*}
in light of (\ref{ny68}), (\ref{ny71}), (\ref{ny80}) and (\ref{ny89}). 
\qed

\begin{rem} \label{ny:rem6} \rm 
In case $u_{2}$ is a unique solution of the equation 
$$
   \frac{\gamma + \delta}{\beta} \psi(u) 
    =  \delta \left( N - \tilde{S}e^{(\beta/\gamma)\tilde{R}}u 
       + \frac{\gamma}{\beta}\log u \right), 
$$
then we get
$$
   T_{2} = \varphi(u_{2}) 
         = \int_{u_{2}}^{e^{-(\beta/\gamma)\tilde{R}}} 
           \frac{d\xi}{\xi\psi(\xi)} 
$$
{\rm ({\it cf}. Figure 6)}. 
If $I'(T_{2}) = 0$, it follows from (\ref{ny3}) that 
$\delta E(T_{2}) = \gamma I(T_{2})$, and 
in Theorem \ref{ny:thm10} we see that 
$$
   \max_{t \geq 0} I(t) = I(T_{2}) = \frac{\delta}{\gamma}E(T_{2}). 
$$
\end{rem}

\begin{figure}[ht]
  \begin{center}
    \includegraphics[height=5cm]{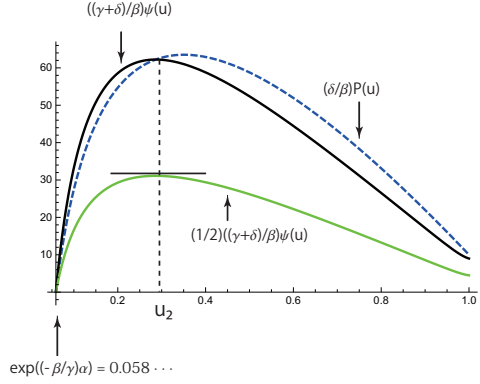}
    \caption{Variations of $(\delta/\beta)P(u) = \delta\bigl(N-\tilde{S}
    e^{(\beta/\gamma)\tilde{R}}u+(\gamma/\beta) \log u)$ (dashed curve), 
    $((\gamma+\delta)/\beta)\psi(u)$ (solid curve) and 
    $(1/2)((\gamma+\delta)/\beta)\psi(u)$ (green curve) 
    obtained by the numerical integration of the initial value problem 
    (\ref{ny90}), (\ref{ny91}) 
    for $N=1000, \tilde{S}=950, \tilde{E}=20, 
    \tilde{I}=30,  \tilde{R}=0, \beta=0.3/1000, \gamma=0.1$ and 
    $\delta=0.2$. In this case we observe that there exists a unique $u_{2}$ 
    such that $(\delta/\beta)P(u_{2}) 
    = ((\gamma+\delta)/\beta)\psi(u_{2})$, and that $T_{2}$ is calculated by }
\end{center}
$$
   T_{2} = \varphi(u_{2}) = \int_{u_{2}}^{1} 
           \frac{d\xi}{\xi\psi(\xi)}, 
$$
where $\psi(u)$ is the unique positive solution of the same initial 
value problem as in Figure 5. 
\end{figure}

\begin{thm} \label{ny:thm11}
The function $E(t) + I(t)$ has the maximum 
$$
   \max_{t \geq 0}\, \bigl(E(t) + I(t)\bigr) 
   = \tilde{S} + \tilde{E} + \tilde{I} 
     - \frac{\gamma}{\beta}\left(1 + \log \tilde{S} 
     - \log \frac{\gamma}{\beta} \right)
$$
at 
$$
   t = T_{3} := \varphi\left( 
   \frac{\gamma}{\beta \tilde{S}e^{(\beta/\gamma)\tilde{R}}} \right) 
   = \int_{\gamma/(\beta \tilde{S}e^{(\beta/\gamma)\tilde{R}})} 
     ^{e^{-(\beta/\gamma)\tilde{R}}} \frac{d\xi}{\xi\psi(\xi)} 
   = S^{-1}\left(\frac{\gamma}{\beta}\right). 
$$
Moreover, $E(t) + I(t)$ is increasing in $[0, T_{3})$ and 
is decreasing in $(T_{3},\infty)$. 
\end{thm}

{\Proof}
From (\ref{ny65}) and (\ref{ny66}) it follows that 
\begin{equation}
   E(t) + I(t) = N - \tilde{S}e^{(\beta/\gamma)\tilde{R}} \varphi^{-1}(t) 
                 + \frac{\gamma}{\beta} \log \varphi^{-1}(t). 
   \label{ny92}
\end{equation}
Differentiating (\ref{ny92}) with respect to $t$ yields 
\begin{eqnarray*}
   E'(t) + I'(t) 
   & = & - \tilde{S}e^{(\beta/\gamma)\tilde{R}} \bigl(\varphi^{-1}(t)\bigr)' 
         + \frac{\gamma}{\beta} 
           \frac{\bigl(\varphi^{-1}(t)\bigr)'}{\varphi^{-1}(t)} \\
   & = & \left( - \tilde{S}e^{(\beta/\gamma)\tilde{R}}\varphi^{-1}(t) 
         + \frac{\gamma}{\beta} \right) 
         \frac{\bigl(\varphi^{-1}(t)\bigr)'}{\varphi^{-1}(t)} \\
   & = & \left( -S(t) + \frac{\gamma}{\beta} \right) 
         \frac{\bigl(\varphi^{-1}(t)\bigr)'}{\varphi^{-1}(t)}. 
\end{eqnarray*}
Since
$$
   \frac{\bigl(\varphi^{-1}(t)\bigr)'}{\varphi^{-1}(t)} 
   = - \psi\bigl(\varphi^{-1}(t)\bigr) < 0 
$$
by (\ref{ny68}), we find that $E'(t) + I'(t) = 0$ for 
$$
   t = T_{3} = \varphi\left( 
   \frac{\gamma}{\beta \tilde{S}e^{(\beta/\gamma)\tilde{R}}} \right) 
   = S^{-1}\left(\frac{\gamma}{\beta}\right). 
$$
We note that 
$$
   e^{-(\beta/\gamma)\alpha}
   < \frac{\gamma}{\beta \tilde{S}e^{(\beta/\gamma)\tilde{R}}} 
   = \frac{\gamma}{\beta \tilde{S}} e^{-(\beta/\gamma)\tilde{R}} 
   < e^{-(\beta/\gamma)\tilde{R}}
$$
by means of (A$_{5}$) and Remark \ref{ny:rem2}. 
In view of (\ref{ny61}) we obtain 
$$
   T_{3} = \varphi\left( 
   \frac{\gamma}{\beta \tilde{S}e^{(\beta/\gamma)\tilde{R}}} \right) 
   = \int_{\gamma/(\beta \tilde{S}e^{(\beta/\gamma)\tilde{R}})} 
     ^{e^{-(\beta/\gamma)\tilde{R}}} \frac{d\xi}{\xi\psi(\xi)} 
   = S^{-1}\left(\frac{\gamma}{\beta}\right).
$$
It is easy to check that $E'(t) + I'(t) > 0$ [resp. $< 0$] 
if and only if $t < T_{3}$ [resp. $> T_{3}$], because 
$\varphi^{-1}(t)$ is decreasing on $[0,\infty)$. 
Therefore we conclude that $E(t) + I(t)$ is increasing in $[0,T_{3})$ and 
is decreasing in $(T_{3},\infty)$. 
It can be shown that 
\begin{eqnarray*}
   \max_{t \geq 0}\, \bigl(E(t) + I(t)\bigr) 
   & = & N - \tilde{S}e^{(\beta/\gamma)\tilde{R}}\varphi^{-1}(T_{3}) 
         + \frac{\gamma}{\beta} \log \varphi^{-1}(T_{3}) \\
   & = & N - \tilde{S}e^{(\beta/\gamma)\tilde{R}}
         \frac{\gamma}{\beta \tilde{S}
            e^{(\beta/\gamma)\tilde{R}}}
         + \frac{\gamma}{\beta}
           \log \left(\frac{\gamma}{\beta \tilde{S}
            e^{(\beta/\gamma)\tilde{R}}}\right) \\
   & = & N - \frac{\gamma}{\beta} 
         + \frac{\gamma}{\beta}\left(\log\, \frac{\gamma}{\beta} 
         - \log\,\tilde{S} - \frac{\beta}{\gamma}\tilde{R}\right) \\
   & = & \tilde{S} + \tilde{E} + \tilde{I}  
         - \frac{\gamma}{\beta}\left(1 + \log\,\tilde{S} 
         - \log\,\frac{\gamma}{\beta}\right). 
\end{eqnarray*}
\qed

\begin{rem} \label{ny:rem7} \rm 
Since $u_{3} = \gamma/\bigl(\beta\tilde{S}e^{(\beta/\gamma)\tilde{R}}\bigr) 
= (\gamma/(\beta\tilde{S}))e^{-(\beta/\gamma)\tilde{R}}$ 
is a unique solution of the equation 
$$ 
   \left( N - \tilde{S}e^{(\beta/\gamma)\tilde{R}}u 
                 + \frac{\gamma}{\beta} \log u \right)' = 0, 
$$
we have 
$$
   T_{3} = \varphi(u_{3}) 
         = \int_{(\gamma/(\beta\tilde{S}))e^{-(\beta/\gamma)\tilde{R}}}^
           {e^{-(\beta/\gamma)\tilde{R}}} 
           \frac{d\xi}{\xi\psi(\xi)} 
$$
{\rm ({\it cf}. Figure 7)}. 
\end{rem}

\begin{figure}[ht]
  \begin{center}
    \includegraphics[height=5cm]{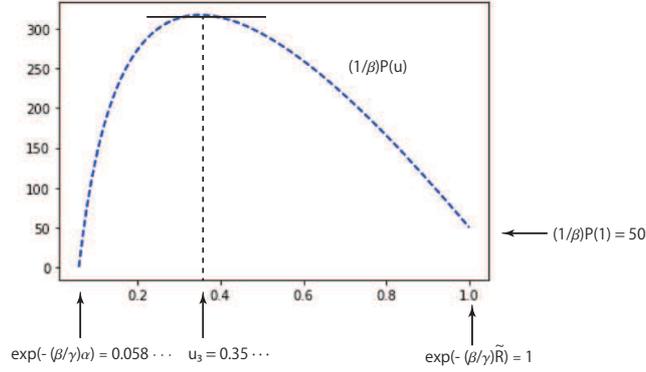}
    \caption{Variation of $(1/\beta)P(u) = N-\tilde{S}
    e^{(\beta/\gamma)\tilde{R}}u+(\gamma/\beta) \log u$ (dashed curve) 
    for $N=1000, \tilde{S}=950, \tilde{E}=20, 
    \tilde{I}=30,  \tilde{R}=0, \beta=0.3/1000, \gamma=0.1$ and 
    $\delta=0.2$. In this case we observe that there exists a unique 
    $u_{3}=20/57=0.35\cdots$ 
    such that $(1/\beta)P'(u_{3})=0$, and that $T_{3}$ is calculated by }
\end{center}
$$
   T_{3} = \varphi(u_{3}) = \int_{20/57}^{1} 
           \frac{d\xi}{\xi\psi(\xi)} = 32.1\cdots, 
$$
where $\psi(u)$ is the unique positive solution of the same initial 
value problem as in Figure 5. 
\end{figure}

\begin{thm} \label{ny:thm12}
The following relation holds{\rm :} 
$$
  S(\infty)  
 = \tilde{S} + \tilde{E} +\tilde{I} 
   + \frac{\gamma}{\beta}\log \frac{S(\infty)}{\tilde{S}}.
$$
\end{thm}

{\Proof}
Since $E(\infty) = I(\infty) = 0$, we deduce that 
\begin{eqnarray*}
   S(\infty)
   & = & N - \tilde{R} + \tilde{R} - R(\infty) \\
   & = & \tilde{S} + \tilde{E} + \tilde{I} 
         + \frac{\gamma}{\beta}\left(\frac{\beta}{\gamma}\tilde{R}
           - \frac{\beta}{\gamma}R(\infty)\right) \\
   & = & \tilde{S} + \tilde{E} + \tilde{I}
         + \frac{\gamma}{\beta}\,\log 
         \left(e^{(\beta/\gamma)\tilde{R}}
         e^{-(\beta/\gamma)R(\infty)}\right) \\
   & = & \tilde{S} + \tilde{E} + \tilde{I}  
         + \frac{\gamma}{\beta}\,\log \frac{S(\infty)}{\tilde{S}} 
\end{eqnarray*}
by means of (\ref{ny88}). 
\qed

\begin{thm} \label{ny:thm13}
We find that 
$$
   S'(\infty) = E'(\infty) = I'(\infty) = R'(\infty) = 0. 
$$
\end{thm}

{\Proof}
Since $E(\infty) = I(\infty) = 0$, we conclude from (\ref{ny1})--(\ref{ny4}) 
that 
\begin{eqnarray*}
   S'(\infty) & = & - \beta S(\infty)I(\infty) = 0, \\
   E'(\infty) & = & \beta S(\infty)I(\infty) - \delta E(\infty) = 0, \\
   I'(\infty) & = & \delta E(\infty) - \gamma I(\infty)  = 0, \\
   R'(\infty) & = & \gamma I(\infty) = 0. 
\end{eqnarray*}
\qed

\begin{rem} \label{ny:rem8} \rm
The hypothesis (A$_{4}$) is satisfied if $\tilde{R} = 0$, 
since $N > \tilde{S}$. 
\end{rem}

\begin{rem} \label{ny:rem9} \rm 
It follows from Theorems \ref{ny:thm7}--\ref{ny:thm10} that 
$S(t) > 0, E(t) > 0, I(t) > 0$ for $t \geq 0$ and $R(t) > 0$ for $t > 0$ 
({\it cf}. Figure 8). 

\end{rem}

\begin{figure}[H]
  \begin{center}
    \includegraphics[height=7cm]{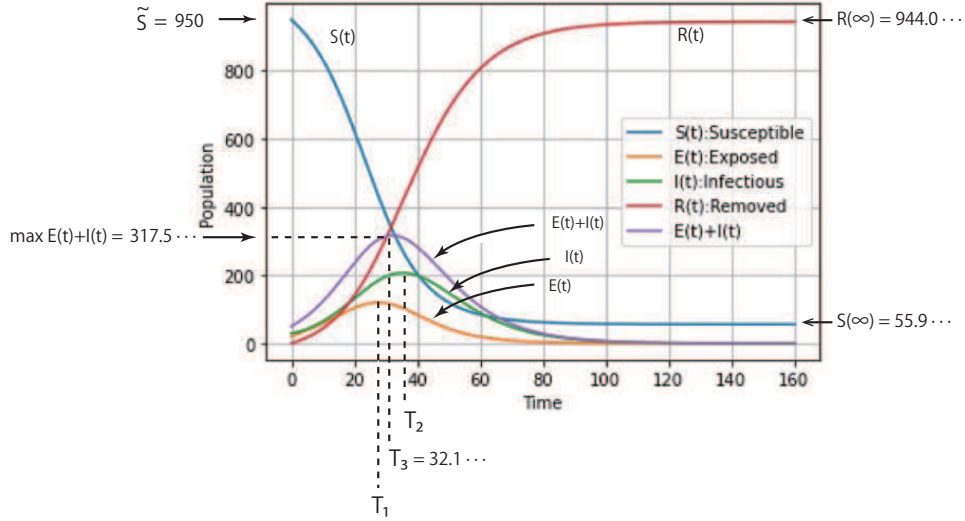}
    \caption{Variations of $S(t)$, $E(t)$, $I(t)$, $R(t)$ and $E(t)+I(t)$ 
obtained by the numerical integration of the initial value problem (\ref{ny1})--(\ref{ny5}) for $N=1000, \tilde{S}=950, \tilde{E}=20, \tilde{I}=30, \tilde{R}=0, \beta=0.3/1000, \gamma=0.1$ and $\delta=0.2$. }
  \end{center}
\end{figure}

\begin{rem} \label{ny:rem10} \rm 
Noting that 
$$
   E_{+}'(0) + I_{+}'(0) = \beta \tilde{S}\tilde{I} - \gamma \tilde{I}, 
$$
we conclude that $E_{+}'(0) + I_{+}'(0) \leq 0$ is equivalent to 
$\tilde{S} \leq \gamma/\beta$.  
We assume that $E_{+}'(0) + I_{+}'(0) \leq 0$. 
Letting $P_{0}(u) := N - \tilde{S}e^{(\beta/\gamma)\tilde{R}}u 
+ (\gamma/\beta)\log u$, we observe that $P_{0}'(u) = 0$ 
at $u = (\gamma/(\beta\tilde{S}))e^{-(\beta/\gamma)\tilde{R}}\ 
(\geq e^{-(\beta/\gamma)\tilde{R}})$ and that $P_{0}(u)$ is 
increasing in 
$\bigl(e^{-(\beta/\gamma)\alpha}, e^{-(\beta/\gamma)\tilde{R}}\bigr]$, 
$\lim_{u \to e^{-(\beta/\gamma)\alpha}+0} P_{0}(u) = 0$ and 
$P_{0}\bigl(e^{-(\beta/\gamma)\tilde{R}}\bigr) = \tilde{E}+\tilde{I}>0$. 
Since $\varphi^{-1}(t)$ is decreasing on $[0,\infty)$, 
$\varphi^{-1}(0) = e^{-(\beta/\gamma)\tilde{R}}$ and 
$\lim_{t \to \infty} \varphi^{-1}(t) = e^{-(\beta/\gamma)\alpha}$, 
it follows that $E(t)+I(t) = P_{0}\bigl(\varphi^{-1}(t)\bigr)$ 
is decreasing on $[0,\infty)$, 
$E(0)+I(0) = \tilde{E}+\tilde{I}$,  
and $E(\infty)+ I(\infty) = 0$ 
{\rm ({\it cf}. {Figure} 9)}. 
\end{rem}

\begin{figure}[H]
  \begin{center}
    \includegraphics[height=5cm]{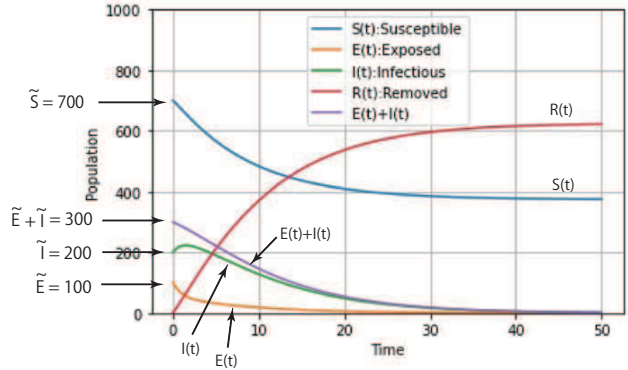}
    \caption{Variations of $S(t)$, $E(t)$, $I(t)$, $R(t)$ and $E(t)+I(t)$ 
obtained by the numerical integration of the initial value problem (\ref{ny1})--(\ref{ny5}) for $N=1000, \tilde{S}=700, \tilde{E}=100, \tilde{I}=200, \tilde{R}=0, \beta=0.2/1000, \gamma=0.2$ and $\delta=0.8$. 
In this case we find that 
$E_{+}'(0) = -52 < 0$, $I_{+}'(0) = 40 > 0$ and 
$E_{+}'(0) + I_{+}'(0) = -12 < 0$. }
  \end{center}
\end{figure}

\noindent
{\bf Acknowledgments} \quad
The author would like to thank Professors Manabu Naito and Hiroyuki Usami 
for their great contributions to the results in Section 3.


\begin{thebibliography}{18}

\bibitem{a65} 
N.~H.~Abel, 
Sur l'\'{e}quation diff\'{e}rentielle $(y+s)dy + (p+qy+ry^{2})dx=0$,
OEuvres compl\`{e}tes de Niels Henrik Abel, S. Lie and L. Sylow, Eds., 
Johnson Reprint Corporation, 
New York, 1965, Vol 2, 26--35. 

\bibitem{b60}
D.~Bernoulli, 
Essai d'une nouvelle analyse de la mortalit\'{e} caus\'{e}e 
par la petite v\'{e}role et des avantages de l'inoculation pour la 
pr\'{e}venir, 
Mem. Math. Phys. Acad. Roy. Sci. (1760), 1--45. 

\bibitem{bst19} M.~Bohner, S.~Streipert and D.~F.~M.~Torres, 
Exact solution to a dynamic SIR model, 
Nonlinear Anal. Hybrid Systems {\bf 32} (2019), 228--238.

\bibitem{bdw08} F.~Brauer, P.~van den Driessche and J.~Wu (Eds.) 
Mathematical Epidemiology, 
Lecture Notes in Mathematics, Vol. 1945, Springer-Verlag, 2008. 

\bibitem{c93} V.~Capasso, 
Mathematical Structures of Epidemic Systems, 
Lecture Notes in Biomathematics, Vol. 97, Springer-Verlag, 1993. 

\bibitem{cl55}
E.~A.~Coddington and N.~Levinson, 
Theory of Ordinary Differential Equations, 
McGraw-Hill, New York, 1955. 

\bibitem{d62} 
H.~T.~Davis, 
Introduction to Nonlinear Differential and Integral Equations, 
Dover Publ., New York, 1962. 

\bibitem{ee04} M.~M.~A.~El-Sheikh and S.~A.~A.~El-Marouf, 
On stability and bifurcation of solutions of an SEIR epidemic model 
with vertical transmission, 
Internat. J. Math. Math. Sci. 2004:56, 2971--2987.

\bibitem{fsaa18} M.~Farman, M.~U.~Saleem, A.~Ahmad and M.~O.~Ahmad, 
Analysis and numerical solution of SEIR epidemic model of 
measles with non-integer time fractional derivatives by using 
Laplace Adomain Decomposition Method, 
Ain Shams Engineering Journal {\bf 9} (2018), 3391-3397.

\bibitem{f80} J.~C.~Frauenthal, 
Mathematical Modeling in Epidemiology, 
Springer-Verlag, Berlin, Heidelberg, 1980. 

\bibitem{hlm14} T.~Harko, F.~S.~N.~Lobo and M.~K.~Mak, 
Exact analytical solutions of the Susceptible-Infected-Recovered 
(SIR) epidemic model and of the SIR model with equal death and 
birth rates,
Appl. Math. Comput. {\bf 236} (2014), 184--194. 

\bibitem{km27} W.~O.~Kermack and A.~G.~McKendrick, 
Contributions to the mathematical theory of epidemics, Part I, 
Proc. Roy. Soc. Lond. Ser. A {\bf 115} (1927), 700--721. 

\bibitem{lm95} M.~Y.~Li and J.~S.~Muldowney, 
Global stability for the SEIR model in epidemiology, 
Math. Biosci. {\bf 125} (1995), 155--164. 

\bibitem{lgwk99} M.~Y.~Li, J.~R.~Graef, L.~Wang and J.~Karsai, 
Global dynamics of a SEIR model with varying total population size, 
Math. Biosci. {\bf 160} (1999), 191--213. 

\bibitem{r76} W.~Rudin, 
Principles of Mathematical Analysis 3rd ed.,
McGraw-Hill, 1976. 

\bibitem{sks10} G.~Shabbir, H.~Khan and M.~A.~Sadiq, 
A note on exact solution of SIR and SIS epidemic models, 
2010, $\langle \mbox{arXiv:1012.5035} \rangle$.

\bibitem{wgk18} J.~Wang, M.~Guo and T.~Kuniya, 
Mathematical analysis for a multi-group SEIR epidemic model with 
age-dependent relapse, 
Appl. Anal. {\bf 97} (2018), 1751--1770. 

\bibitem{whkb20} S.~J.~Weinstein, M.~S.~Holland, K.~E.~Rogers 
and N.~S.~Barlow, 
Analytic solution of the SEIR epidemic model via asymptotic approximant, 
Physica D {\bf 411} (2020) 132633. 

\end{thebibliography}
\end{document}